\documentclass[twocolumn,twocolappendix]{aastex631}
\usepackage{multirow}

\newcommand{\target}{V1298\,Tau}
\newcommand{\xmm}{{\em XMM-Newton}}

\newcommand{\fxu}{{erg\,s$^{-1}$\,cm$^{-2}$}}

\newcommand{\lxu}{{erg\,s$^{-1}$}}

\received{March 9, 2023}
\accepted{May 5, 2023}

\shorttitle{XUV emission of V1298\,Tau}
\shortauthors{A. Maggio et al.}
\graphicspath{{./}{figures/}}

\begin{document}

\title{XUV emission of the young planet-hosting star V1298\,Tau from coordinated observations with XMM-Newton and HST
}

\author[0000-0001-5154-6108]{A. Maggio}
\affiliation{INAF -- Osservatorio Astronomico di Palermo, Piazza del Parlamento, 1, I-90134, Palermo, Italy}

\author[0000-0003-4948-6550]{I. Pillitteri}
\affiliation{INAF -- Osservatorio Astronomico di Palermo, Piazza del Parlamento, 1, I-90134, Palermo, Italy}

\author[0000-0003-2073-1348]{C. Argiroffi}
\affiliation{Department of Physics and Chemistry, University of Palermo, Piazza del Parlamento, 1 I-90134, Palermo, Italy}
\affiliation{INAF -- Osservatorio Astronomico di Palermo, Piazza del Parlamento, 1, I-90134, Palermo, Italy}

\author[0000-0002-4638-3495]{S. Benatti}
\affiliation{INAF -- Osservatorio Astronomico di Palermo, Piazza del Parlamento, 1, I-90134, Palermo, Italy}

\author[0000-0002-1600-7835]{J. Sanz-Forcada}
\affiliation{Centro de Astrobiolog\'{i}a (CSIC-INTA), ESAC Campus, E-28692 Villanueva de la Ca\~nada, Madrid, Spain}

\author[0000-0002-2662-3762]{V. D'Orazi}
\affiliation{INAF -- Osservatorio Astronomico di Padova, Vicolo dell’Osservatorio 5, I-35122 Padova, Italy}
\affiliation{Department of Physics, University of Rome "Tor Vergata", Via della Ricerca Scientifica 1, I-00133 Roma, Italy}

\author[0000-0002-1892-2180]{K. Biazzo}
\affiliation{INAF–Osservatorio Astronomico di Roma, Via Frascati 33, I-00040 Monte Porzio Catone, Italy}

\author[0000-0003-4830-0590]{F. Borsa}
\affiliation{INAF– Osservatorio Astronomico di Brera, Via E. Bianchi 46, I-23807 Merate, Italy}

\author[0000-0002-5130-4827]{L. Cabona}
\affiliation{INAF -- Osservatorio Astronomico di Padova, Vicolo dell’Osservatorio 5, I-35122 Padova, Italy}

\author[0000-0001-7707-5105]{R. Claudi}
\affiliation{INAF -- Osservatorio Astronomico di Padova, Vicolo dell’Osservatorio 5, I-35122 Padova, Italy}

\author[0000-0001-8613-2589]{S. Desidera}
\affiliation{INAF -- Osservatorio Astronomico di Padova, Vicolo dell’Osservatorio 5, I-35122 Padova, Italy}

\author[0000-0002-9824-2336]{D. Locci}
\affiliation{INAF -- Osservatorio Astronomico di Palermo, Piazza del Parlamento, 1, I-90134, Palermo, Italy}

\author[0000-0003-1149-3659]{D. Nardiello}
\affiliation{INAF -- Osservatorio Astronomico di Padova, Vicolo dell’Osservatorio 5, I-35122 Padova, Italy}

\author[0000-0002-9428-8732]{L. Mancini}
\affiliation{Department of Physics,
University of Rome "Tor Vergata",
Via della Ricerca Scientifica 1,
I-00133 Roma, Italy}
\affiliation{Max Planck Institute for Astronomy, K\"{o}nigstuhl 17,
D-69117, Heidelberg, Germany}
\affiliation{INAF -- Osservatorio Astrofisico di Torino, via Osservatorio 20, I-10025, Pino Torinese, Italy}

\author[0000-0002-9900-4751]{G. Micela}
\affiliation{INAF -- Osservatorio Astronomico di Palermo, Piazza del Parlamento, 1, I-90134, Palermo, Italy}

\author[0000-0002-8786-2572]{M. Rainer}
\affiliation{INAF– Osservatorio Astronomico di Brera, Via E. Bianchi 46, 23807 Merate, Italy}

\author[0000-0001-5252-5042]{R. Spinelli}
\affiliation{INAF– Osservatorio Astronomico di Brera, Via E. Bianchi 46, 23807 Merate, Italy}

\author[0000-0002-5606-6354]{A. Bignamini}
\affiliation{INAF – Osservatorio Astronomico di Trieste, via Tiepolo 11, I-34143 Trieste, Italy}


\author[0000-0001-9984-4278]{M. Damasso}
\affiliation{INAF – Osservatorio Astrofisico di Torino, Via Osservatorio 20, 10025 Pino Torinese, Italy}

\affiliation{INAF -- Osservatorio Astronomico di Padova, Vicolo dell’Osservatorio 5, I-35122 Padova, Italy}


\begin{abstract}
Atmospheric mass loss plays a major role in the evolution of exoplanets. This process is driven by the stellar high-energy irradiation, especially in the first hundreds of millions of years after dissipation of the proto-planetary disk. A major source of uncertainty in modeling atmospheric photo-evaporation and photo-chemistry is due to the lack of direct measurements of the stellar flux at EUV wavelengths. Several empirical relationships have been proposed in the past to link EUV fluxes to emission levels in X-rays, but stellar samples employed for this aim are heterogeneous, and available scaling laws provide significantly different predictions, especially for very active stars.
We present new UV and X-ray observations of \target\ with HST/COS and XMM-Newton, aimed to determine more accurately the XUV emission of this solar-mass pre-Main Sequence star, which hosts four exoplanets. Spectroscopic data were employed to derive the plasma emission measure distribution vs.\ temperature, from the chromosphere to the corona, and the possible variability of this irradiation  on short and year-long time scales, due to magnetic activity. As a side result, we have also measured the chemical abundances of several elements in the outer atmosphere of \target.
We employ our results as a new benchmark point for the calibration of the X-ray to EUV scaling laws, and hence to predict the time evolution of the irradiation in the EUV band, and its effect on the evaporation of exo-atmospheres.
\end{abstract}



\section{Introduction} \label{sec:intro}

The frequency of planets as a function of their masses, size, and host star properties is a key parameter for testing planet formation and evolution models.
On the other hand, evolutionary paths are the result of a complex interplay between physical and dynamic processes operating on different time scales, including the stellar
radiation fields. In particular, intense high-energy irradiation from the host stars, especially at young ages, can be responsible for atmospheric evaporation of the exoplanets, and it is one of the ingredients, still poorly understood, that shapes the planet mass-radius relationship (\citealt{Lopez+Fortney2013, Ower+Wu2013, Fulton+2017, Owen+Wu2017, Fulton+Petigura2018, Owen+Lai2018}).

The thermal structure and chemistry of planetary atmospheres sensitively depend on the spectral energy distribution of the stellar radiation \citep{Lammer+2003}. While EUV photons are absorbed in the upper atmosphere, soft X-rays can heat and ionize lower layers due to secondary electron production \citep{ChecchiPestellini+2006}. 
A reliable characterization of planetary evolution requires the knowledge of the stellar XUV emission (5--920\,\AA\ range) and its variation with stellar age (\citealt{SF11,Locci+2019}).

Planets in close orbits around young stars are especially susceptible to the effects of irradiation because the host stars are known to have higher magnetic activity levels relative to the Sun.
Higher XUV fluxes are generally accompanied by more frequent and energetic flares, and conjectured higher rates of Coronal Mass Ejections (CMEs, \citealt{Khodachenko+2007}).
In turn, charged particle flows linked to stellar winds and CMEs determine the size and time-dependent compression of planetary magnetospheres, and eventually may lead to stripping (erosion) of close-in planets \citep{Lammer+2007}, as well as deposition of gravity waves \citep{Cohen+2014}.


High energy radiation (including X-rays, Extreme UV, and Far UV bands)
originates from stellar outer atmospheres. In particular, plasmas in the temeprature range 
T$\sim10^4-10^5.5$ K produces FUV emission lines, while X-ray spectral lines and continuum emission 
are tipically formed where temperatures rise above $10^6-10^7$ K. 
Significant flux in the EUV band originates from plasma with temperatures in the whole range $10^4-10^{7.4}$\,K.

The goal of coordinated observations at X-rays and UV wavelengths is to provide a benchmark for models of photo-evaporation and photo-chemistry of exoplanets, especially at young ages.
Measurements of the hardness, fluence, and variability of the XUV irradiation, can be connected to masses, sizes and atmospheric chemical composition, acquired with observations from ground and space-borne facilities. 


Here we present a detailed study of the young star \target, simultaneously observed at UV wavelengths with HST/COS and in X-rays with XMM-Newton.


This target represents a unique study case, because it is a young solar analogue hosting a compact planetary system, with four planets at distances between 0.08 and 0.3\,AU from the host star, which implies more than a factor 10 difference in XUV irradiation. Hence, the system provides a valuable benchmark for planet migration and evolution models, and rich prospects for atmospheric characterization by transmission spectroscopy.

In this paper we introduce first the characteristics of the host star and its planetary system (Sect. \ref{sect:star}). Next, we present new HST/COS and \xmm\ observations of the host star (Sect. \ref{sec:xuv}), performed simultaneously, with the aim to reconstruct the stellar spectrum over a wide wavelength range, extending from 5~\AA\ to 1450~\AA.
Then, we describe our reconstruction of the transition region and coronal plasma distribution vs.\ temperature, and revise the most likely radiation budget at EUV wavelengths (Sect. \ref{sect:dem}). Finally, we compare our results to predictions based on X/EUV scaling laws proposed in the past, and we draw attention on their accuracy and limitations (Sect. \ref{sec:discuss}).

\section{Stellar characteristics}
\label{sect:star}
 \target\ is a K1 star with a mass of $1.170\pm0.060$\,M$_{\odot}$, a radius of $1.278\pm0.070$\,R$_{\odot}$, an effective temperature $T_{\rm eff} = 5050\pm100$\,K, and a bolometric luminosity $L_{\rm bol} = 0.954\pm0.040$\,L$_{\odot}$ \citep{paperI+2021}. 
 It is located at a distance of $108.02\pm0.23$\,pc \citep[DR3]{GaiaDR3} toward the Taurus region, and it belongs to the Group 29 stellar association \citep{Oh+2017}. For \target\ we determined an age of $11.9^{+2.0}_{-3.2}$\,Myr \citep[]{Maggio+2022}.


\target\ hosts one of the youngest planetary systems known so far, discovered with the 
Kepler K2 mission \citep{David+2019b}.
The system counts two Neptune-sized planets (dubbed "c" and "d"), one Jovian planet ("b"),
and one Saturn-sized planet ("e"), in order of increasing separation from the central 
star \citep{paperI+2021}.

Possible alternative evolutionary paths of the planetary masses and sizes, due to photoevaporation of the primary atmospheres, were explored by \cite{Poppenhaeger+2020}, based on a X-ray snapshot with Chandra, and by \cite{Maggio+2022}, based on a first observation of \target\ with XMM-Newton, performed in February 2021.

\begin{figure}[!t]
\begin{center}
\resizebox{\columnwidth}{!}{ 
\includegraphics{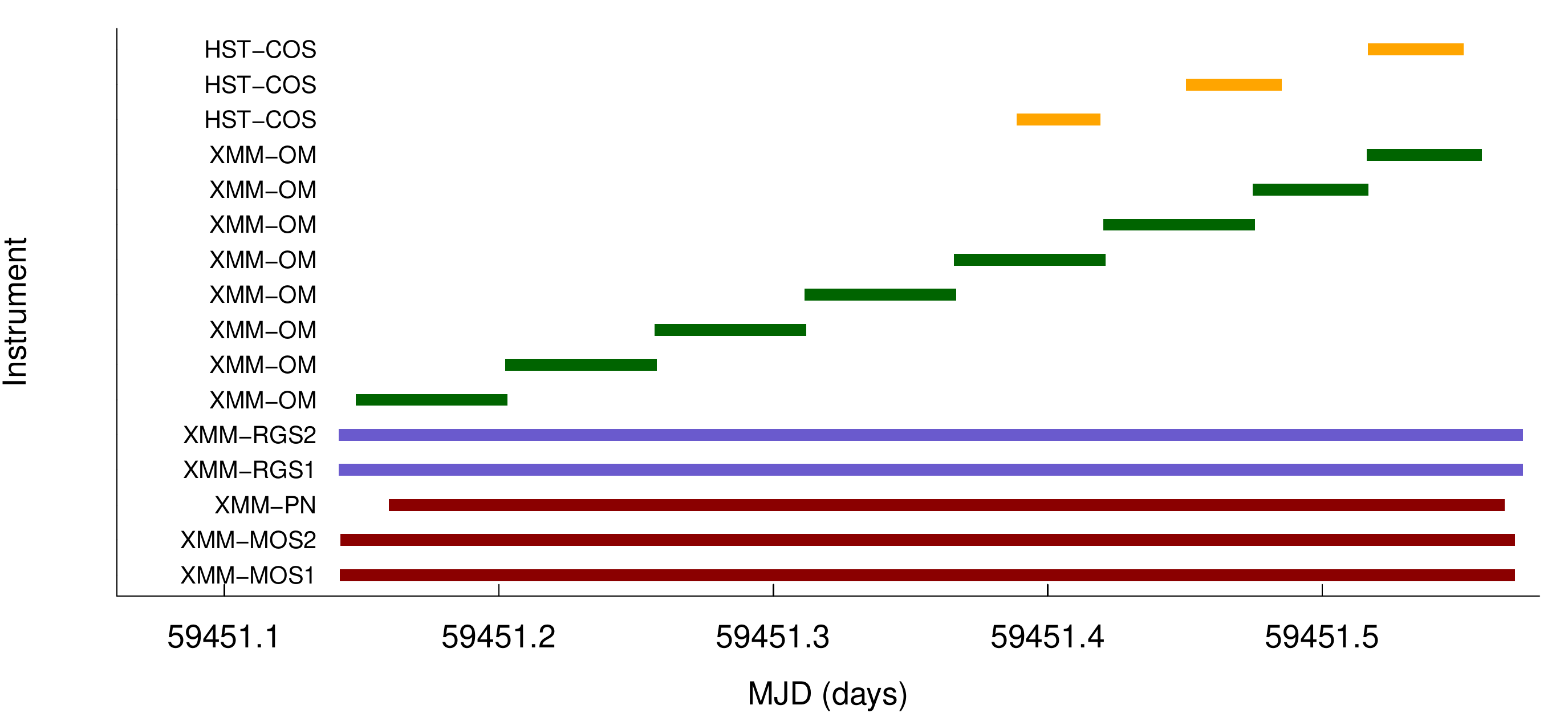} }
\end{center}
\caption{\xmm\ and HST timing of the observations. The colored bars represent
different instruments (EPIC, RGS, OM, and COS). HST covered the second part
of the \xmm\ exposure which started about 20 ks before the first HST visit.}
\label{fig:mjd}
\end{figure}

\begin{figure}[!t]
\begin{center}
\resizebox{\columnwidth}{!}{ 
\includegraphics[width=\hsize]{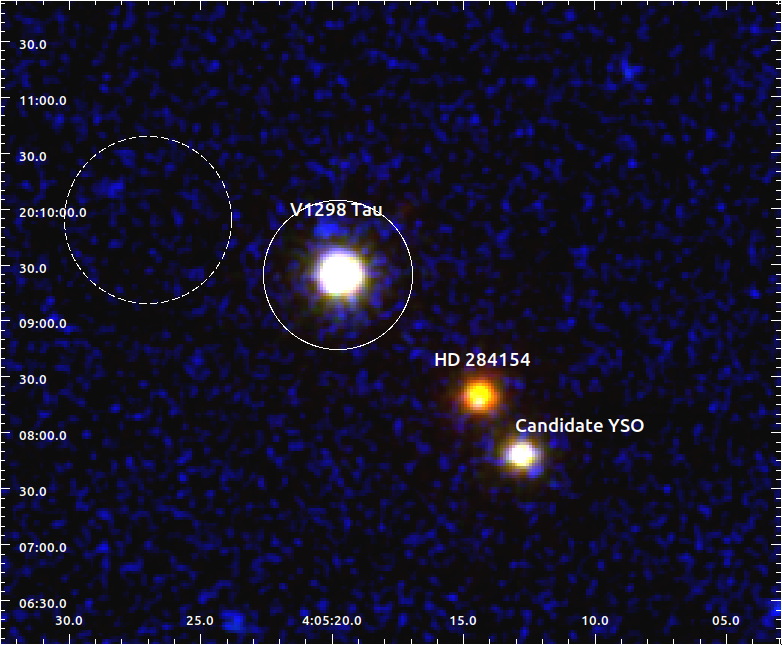} }
\end{center}
\caption{\xmm\ EPIC RGB image of \target. Red: $0.3-1.0$ keV, Green: $1.0-3.0$ keV, Blue: $3.0-8.0$ keV. The target and the two bright nearby stars are labelled. 
The regions used for accumulating the spectra of \target\ and the background are marked with white circles.\label{fig:x} }
\end{figure}

   
\begin{figure}[t]
\begin{center}
\resizebox{\columnwidth}{!}{\includegraphics{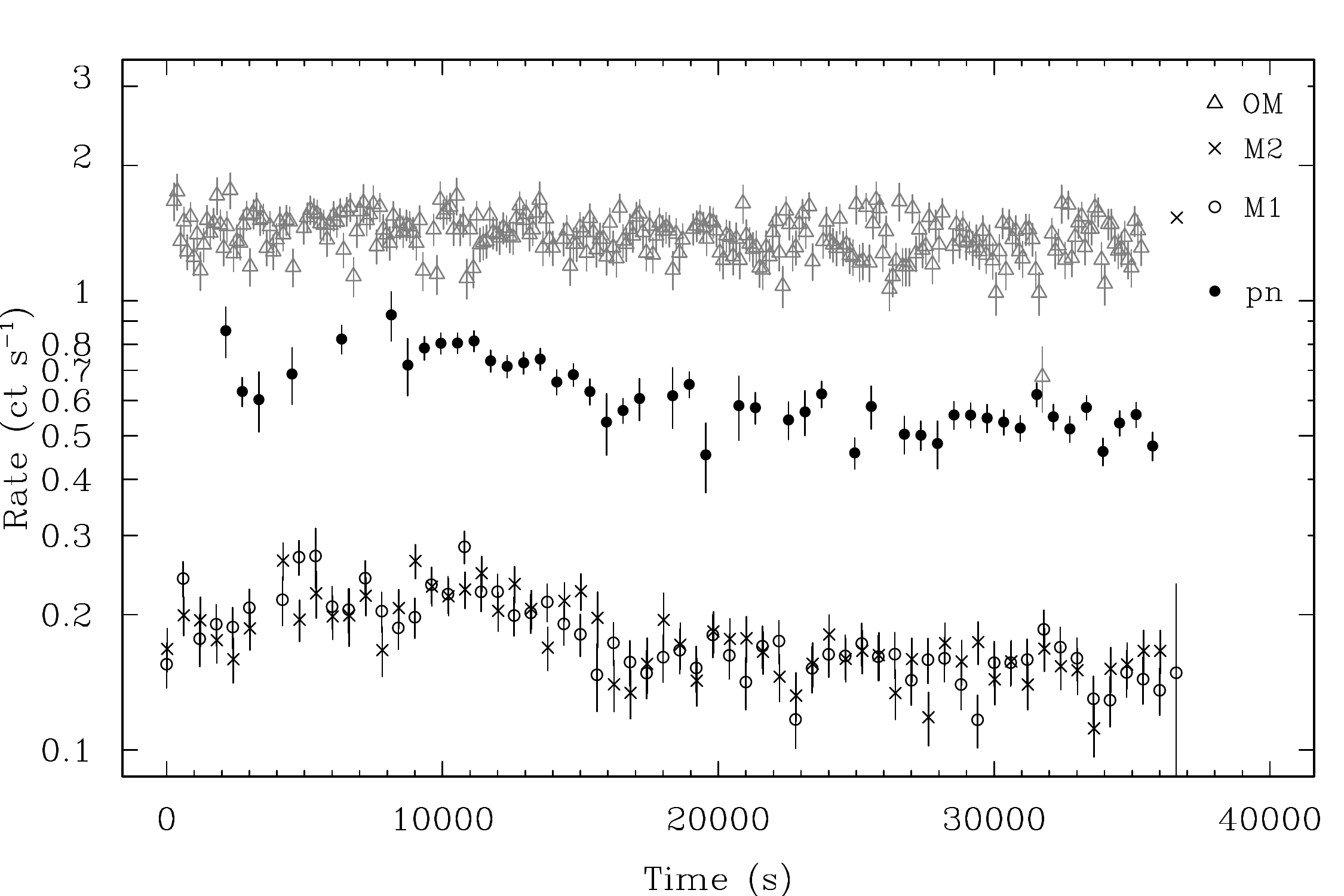} }
\resizebox{\columnwidth}{!}{\includegraphics[trim= 3.5cm 1.6cm 4.8cm 1cm, clip]{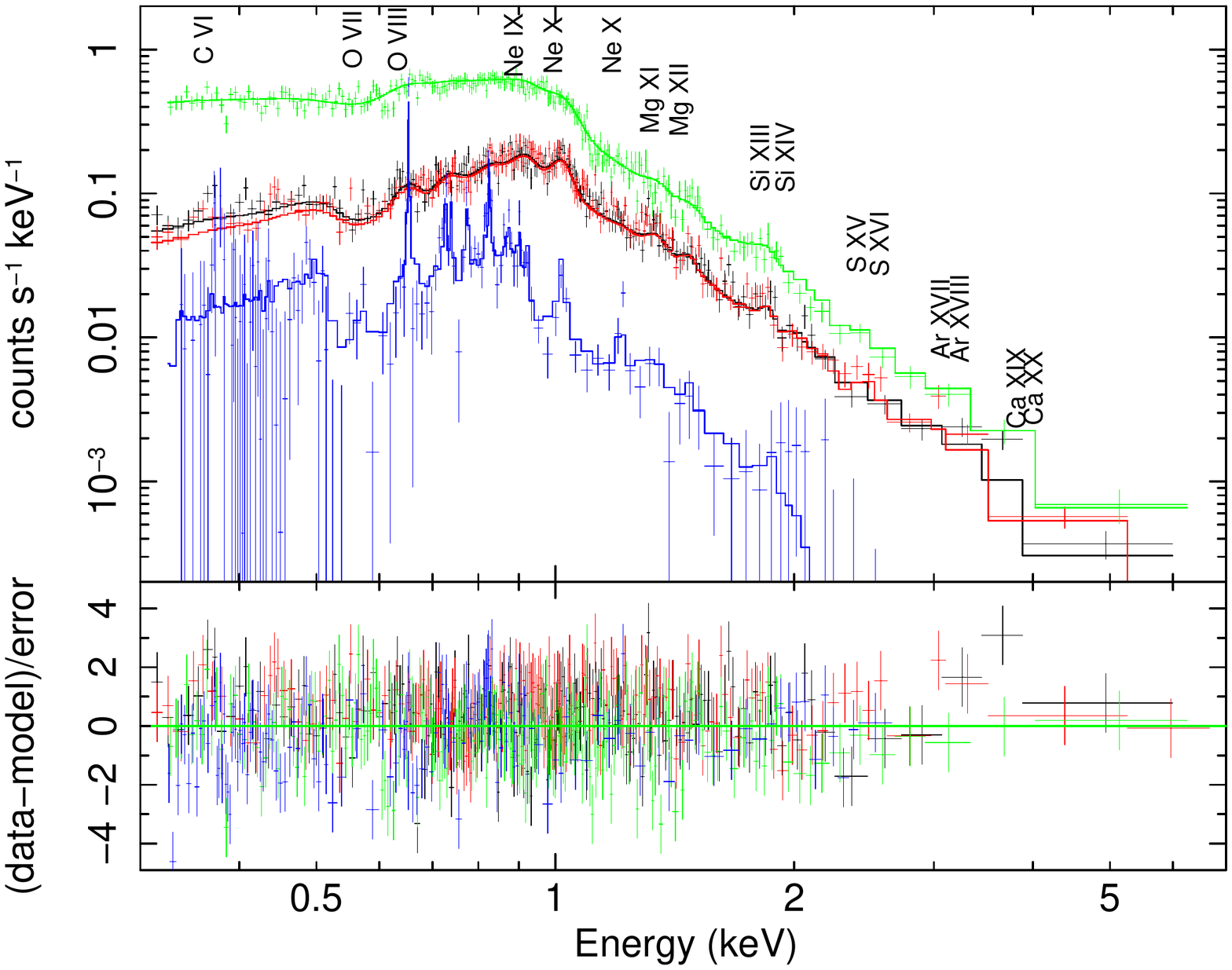} }

\end{center}
\caption{\footnotesize{Top panel: pn, MOS (time bin size of 600\,s) and OM light curves (time bin size of 120\,s) during the Aug 2021 observation. 
Bottom panel: EPIC and RGS spectra of \target\ with best-fitting model and residuals for the full data set (pn in green, MOS1/MOS2 in red/black, RGS in blue). Labels indicate the location of prominent emission lines of H-like and He-like ions.
\label{fig:xraylc}   }
}
 \end{figure}

\section{Observations and data analysis}
\label{sec:xuv}
\subsection{XMM-Newton observation}
\label{sec:xmm}

A new \xmm\ observation of \target\ was performed on 2021 August 25 (ObsId 0881220101, PI S.\,Benatti), in order to characterize the X-ray emission and variability of the host star. 
The exposure time was about 36\,ks with EPIC as a prime instrument in {\it full frame} window imaging mode and with the {\it Medium} filter. 
We also acquired data from RGS and OM instruments simultaneously.

The Observation Data Files (ODFs) were reduced with the Science Analysis System (SAS, ver.20.0.0), following standard procedures. 
We obtained FITS tables of X-ray events detected by the three EPIC CCD cameras (MOS1, MOS2, and pn) and with the two high-resolution spectrographs (RGS1 and RGS2), calibrated in energy, arrival times and astrometry by means of the {\it emchain/epchain} SAS tasks. 
Inspection of the light curve of events detected with energies $> 10$\,keV allowed us to identify and filter out several time intervals affected by high background. 


\target\ is sufficiently isolated to allow the extraction of the source signal from
a circular region of 40\arcsec radii for MOS and pn, and local background from an uncontaminated nearby circular region of similar size. With SAS we also produced the response matrices and effective area files needed for the subsequent spectral analysis. 
RGS source and background spectra were also extracted adopting the standard results of the SAS pipeline. Source X-ray spectra and light curves are shown in Fig. \ref{fig:xraylc}. In the same figure, we show the OM light curve, obtained with the SAS task {\sc omfchain}.


\begin{figure*}
    \centering
    \resizebox{1.0\textwidth}{!}{\includegraphics{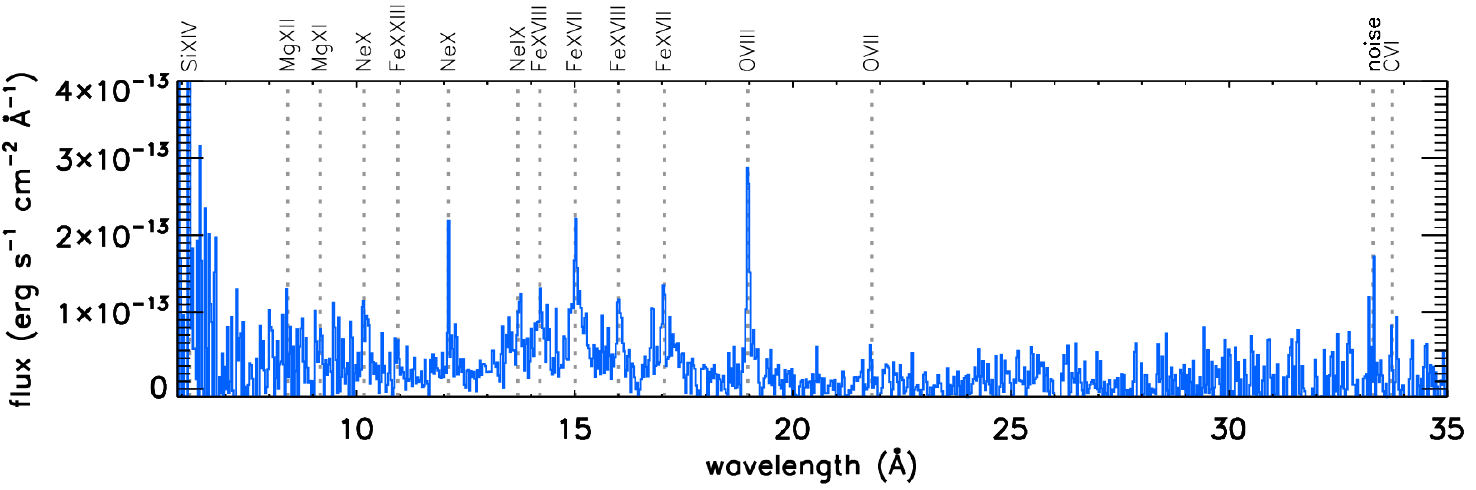}}
    \caption{High-resolution X-ray spectrum of \target\ observed with RGS, rebinned by a factor of 3, with labels marking some of the strongest emission lines. We have also labelled as "noise" a spurious peak near the \ion{C}{6} line at 33.7 \AA. }
    \label{fig:rgsspec}
\end{figure*}

\begin{figure}
    \centering
    \resizebox{0.49\textwidth}{!}{\includegraphics{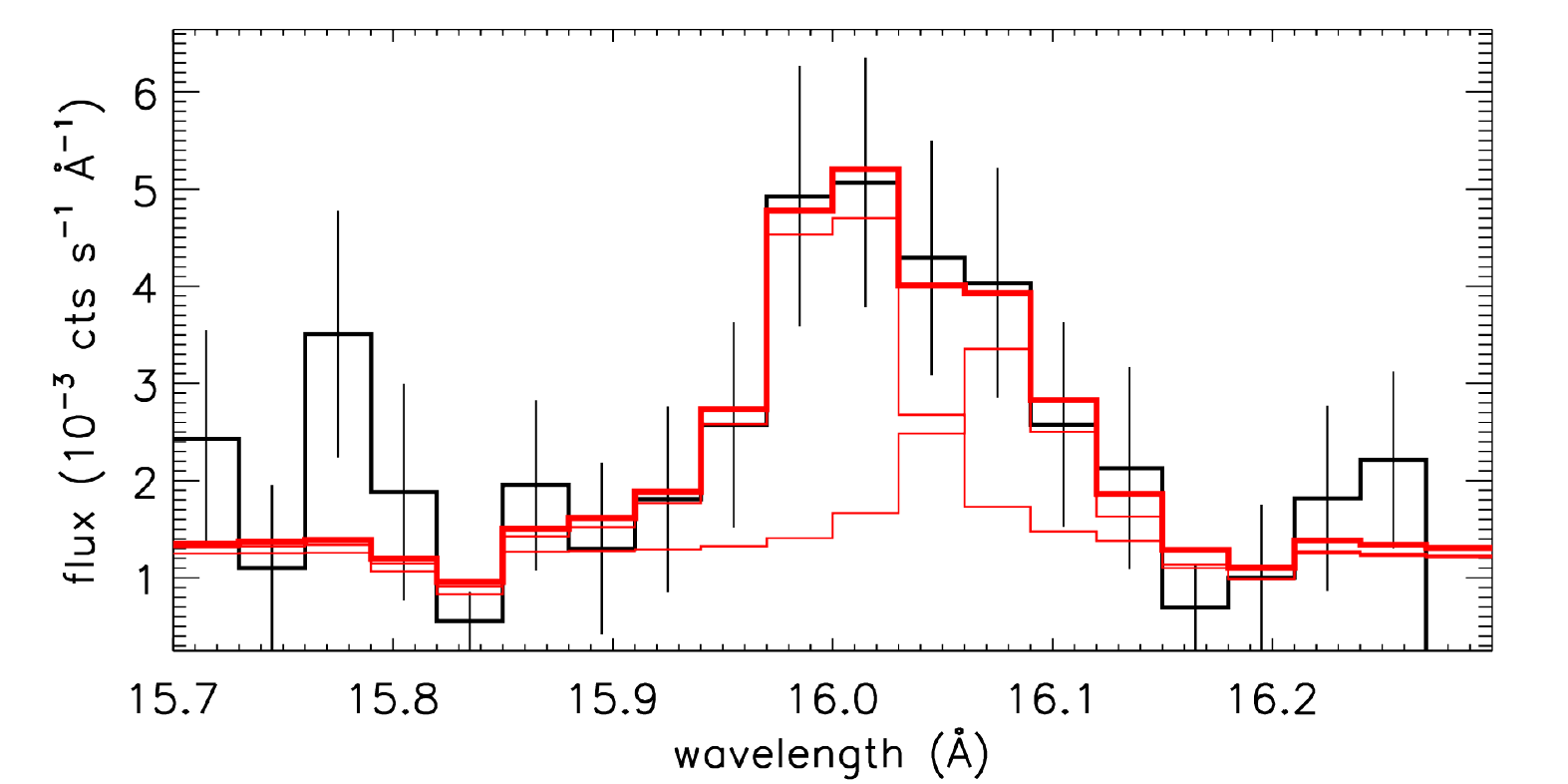}}
    \resizebox{0.49\textwidth}{!}{\includegraphics{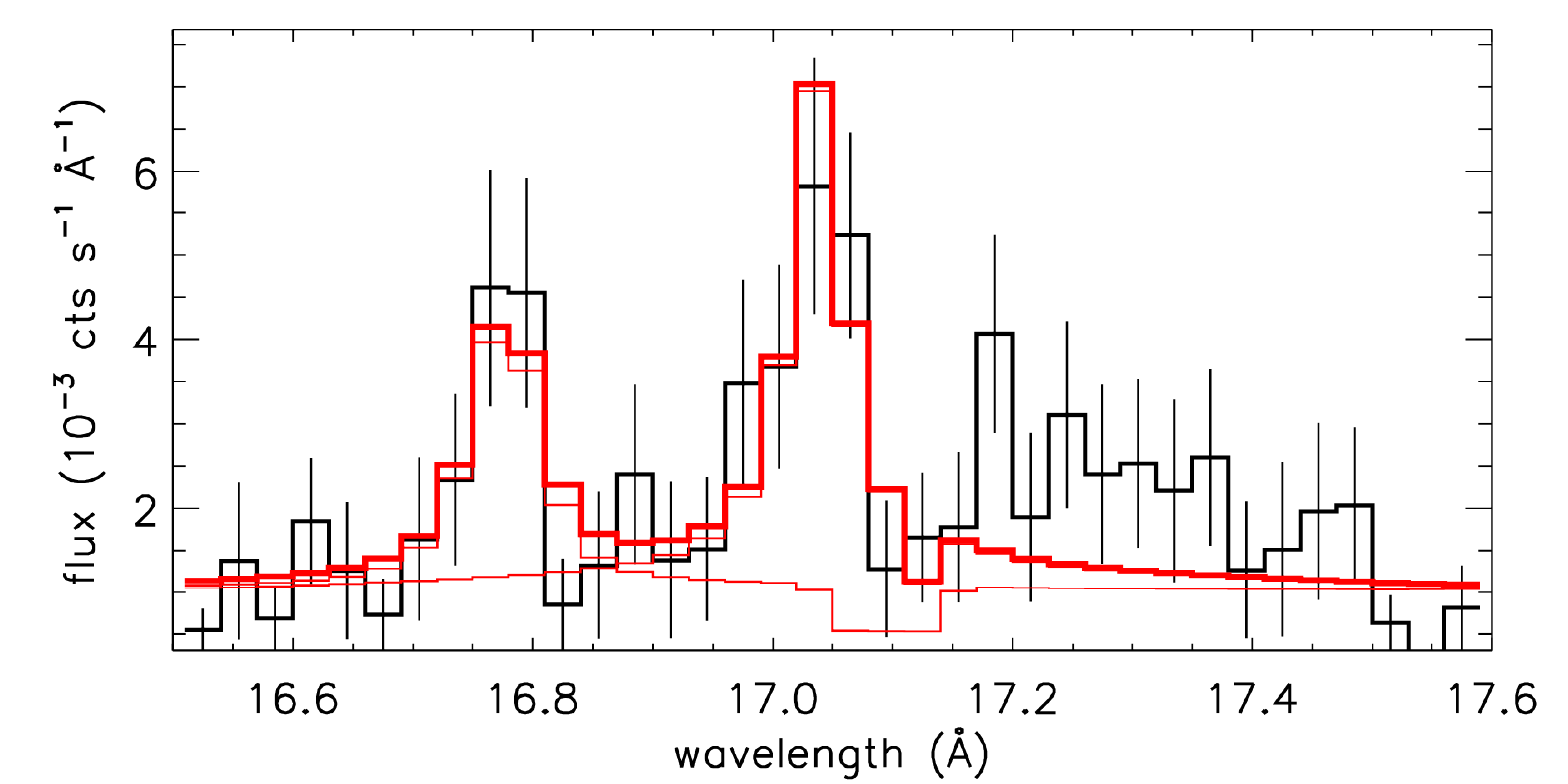}}
    \caption{Examples of RGS line fitting. The 1-st order RGS spectrum, in black, in two small wavelength regions, with superimposed the best-fit function used to infer line fluxes. Thick red line marks the total best-fitting function, thin red line the contribution of individual emission lines.}
    \label{fig:rgslinefit}
\end{figure}

For measurements of X-ray line fluxes we considered the spectrum obtained by adding the 1-st order RGS1 and RGS2 spectra, which cover the $5-40$\,\AA\, range, with an average resolution of 0.06\,\AA\, FWHM. For improving the statistics, we also co-added the RGS spectra obtained in Aug 2021 with those already available from the previous XMM-Newton observation, taken in Feb 2021 \citep[]{Maggio+2022} ($\sim 27$\,ks of clean exposure time), and we rebinned the resulting spectrum by a factor of 3 (Fig.~\ref{fig:rgsspec}). This choice is supported by the lack of strong variability of the coronal emission (Sect.\ref{sect:xspec}).

We identified the strongest emission lines and measured their fluxes by fitting the observed spectrum in small wavelength intervals. To model the line profile we used the RGS line spread function tabulated in the RGS response matrix file. The width of each wavelength interval was set to include blended lines that needed simultaneous fitting. In the best-fitting function, in addition to individual line contributions, we included also a continuum component, with temperature and normalization left free to vary. Because of the small width of the selected wavelength intervals, the continuum component, acting as an additive constant, takes into account also the integrated contribution of unresolved weak lines. Two examples of RGS line fitting are plotted in Fig.~\ref{fig:rgslinefit}. The measured line fluxes are listed in Table~\ref{tab:linefluxes}. The observed X-ray lines form in the temperature range $\sim 1$--16\,MK, probing the thermal structure of the stellar corona.

\subsection{HST observations}
\label{sec:hst}

\begin{figure*}
    \centering
    \resizebox{1.0\textwidth}{!}{\includegraphics{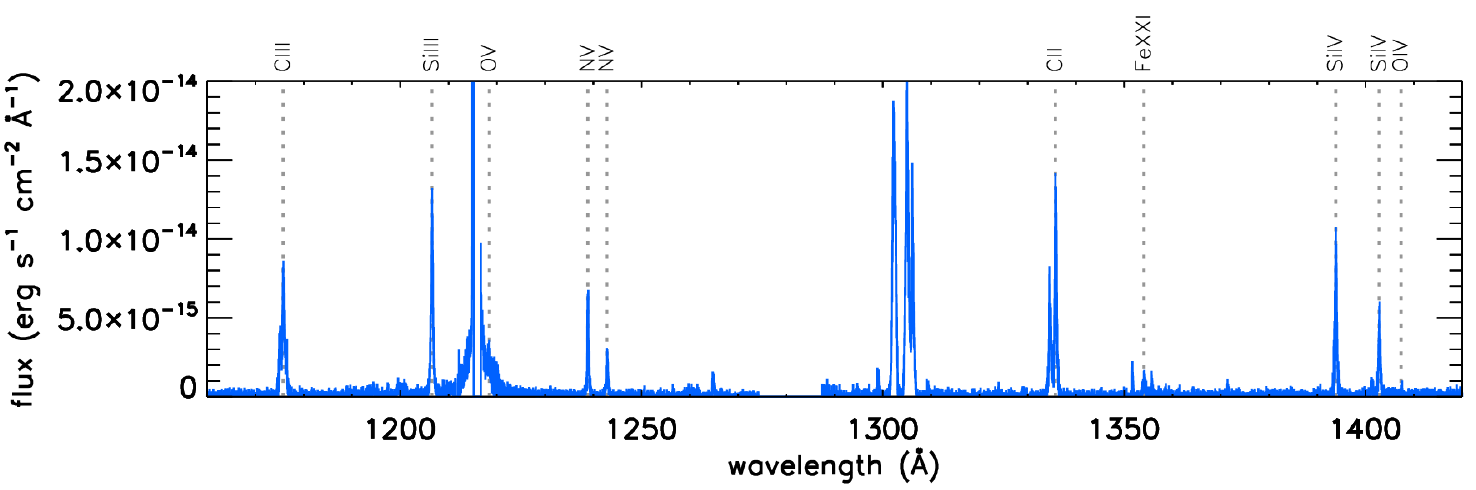}}
    \caption{\target\ high-resolution UV spectrum observed with COS G130.}
    \label{fig:spec_cos}
\end{figure*}

\target\ was observed with HST during three consecutive orbits to acquire spectra in FUV with the COS spectrograph. The central wavelength was 1291\AA\ covering about the 1150 -- 1450 \AA\ wavelength range. About 55 minutes were available during each orbit for science and acquisition exposures of \target. The actual exposure times were 2249.2\,s, 2634.1\,s, and 2634.2\,s, respectively, during the three orbits totalling about 7517.5\,s. 
The orbits were simultaneous with the second half of the \xmm\  observation during which the star appears quiescent in X-rays. The standard calibration operated with CalCOS (v.\ 3.3.10) at STScI was deemed sufficient and thus the reduced spectra were downloaded from the HST archive and ready for the subsequent analysis. The archive provides the spectra accumulated during each orbit and their average spectrum. We checked that the ion lines have similar intensities during each orbit. Since the star was quiescent in X-rays, we determined to analyze the average spectrum (Fig.\ \ref{fig:spec_cos}) so to maximize the count statistics.

We identified and measured the strongest UV emission lines in this averaged spectrum. Line fluxes were obtained by fitting the observed spectrum in small wavelength intervals, assuming for the line profile a beta-model (also known as Moffat line profile)

\begin{figure}
    \centering
    \resizebox{0.47\textwidth}{!}{\includegraphics{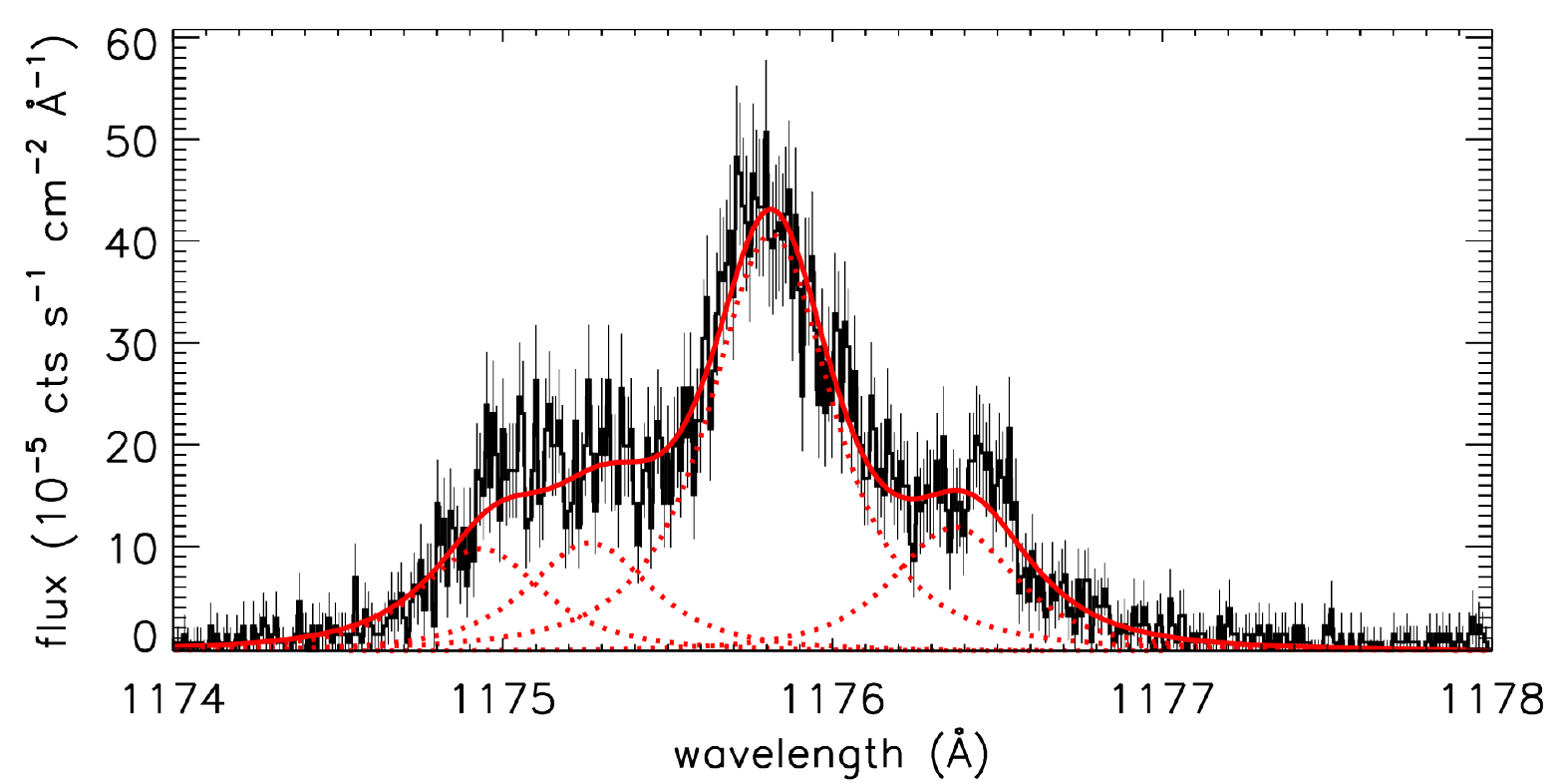}}
    \resizebox{0.47\textwidth}{!}{\includegraphics{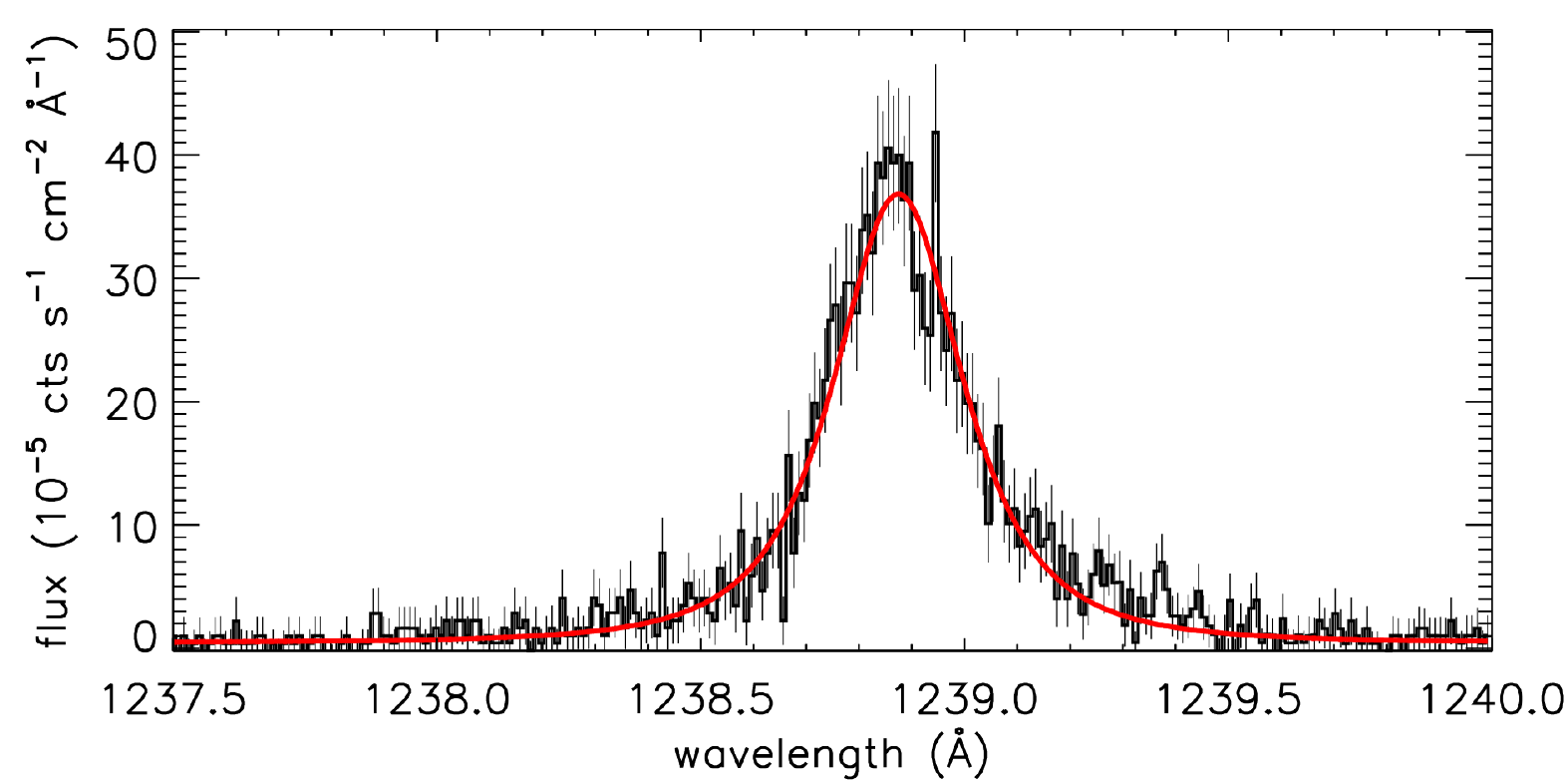}}
    \caption{Examples of line fitting of the COS spectrum (in black) in two small wavelength regions, with superimposed contributions of individual emission lines (dotted red lines) used to infer line fluxes, and the total best-fit function (solid red line). The top panel shows the case of the \ion{C}{3} ion, with six contributing lines, but only four resolved, while a \ion{N}{5} line is shown in the bottom panel.}
    \label{fig:coslinefit}
\end{figure}

\begin{equation}
f\left(\lambda\right)=f_{\mathrm{max}}\left[1+\left(\frac{\lambda- \lambda_{\mathrm{c}}}{\Delta\lambda}\right)^{2}\right]^{-\beta}
\end{equation}
with $\Delta\lambda=0.03$\,\AA\ and $\beta=1.6$ kept as fixed parameters. We adopted this line shape and parameters after having checked that this function reproduces well the observed line profile of strong and isolated lines. 
The width of the selected wavelength intervals was set to eventually include blended lines, for which simultaneous fitting is needed. We also included a constant function in addition to the line contributions in each interval, to fit also the continuum level. Two examples of COS line fitting are shown in Fig.~\ref{fig:coslinefit}, and the case of a few lines possibly altered by opacity effects are described in Appendix \ref{Appendix}. 

Besides our Aug 2021 HST/COS G130M observation, we have also considered a previous observation of \target\ taken with the G160M grating on 2020 Oct 17 (ObsId lebw02koq, PI P.W.\ Cauley), and available in the MAST archive. The aim was to measure the flux of the \ion{C}{4} doublet at $\sim1550$\,\AA, which is important to better constrain the thermal structure of the stellar chromosphere (Sect.\ \ref{sect:dem}). Since this observation was acquired in a different epoch, we checked again for the presence of variability by comparing the fluxes of the \ion{Si}{4} doublet at $\sim1400$\,\AA\, which is present in both the 2021 and 2020 spectra. As a result, we corrected this effect by dividing the \ion{C}{4} flux by a factor of 2.7.

All the measured line fluxes are listed in Table~\ref{tab:linefluxes}. This list comprises UV lines that form in the temperature range $\sim 30,000-200,000$\,K, thus probing the thermal structure of the high chromosphere and the transition region of the star. In the same list we reported also the fluxes of a couple of lines with no clear identification. We instead did not include some \ion{O}{1} emission lines, usually present in COS spectra, because of their geocoronal contamination.

\begin{table*}
\caption{Measured X-ray and UV line fluxes of V1298~Tau.}
\label{tab:linefluxes}
\scriptsize
\begin{center}
\begin{tabular}{rlcr@{$\;\pm\;$}lcccc}
\hline\hline
\multicolumn{1}{c}{$\lambda^{a}$} & Ion$^{b}$ & $\log T_{\rm max}^{c}$ & \multicolumn{2}{c}{$F_{\rm obs}^{d}$} & EMD1$^e$ & $((F_{\rm obs}-F_{\rm pred})/\sigma)^f$ & EMD2$^e$ & $((F_{\rm obs}-F_{\rm pred})/\sigma)^g$ \\
\hline
    6.18 &                                                   \ion{Si}{14}\,\ion{Si}{14}\, &  7.20 &       610 &       250 &        &             & $\ast$ &        1.43 \\
    8.42 &                                                   \ion{Mg}{12}\,\ion{Mg}{12}\, &  7.00 &        56 &        49 & $\ast$ &       -0.39 & $\ast$ &       -0.56 \\
    9.17 &                                                                   \ion{Mg}{11}\, &  6.80 &        64 &        41 & $\ast$ &        0.41 & $\ast$ &        0.30 \\
   10.24 &                                                       \ion{Ne}{10}\,\ion{Ne}{10}\, &  6.80 &       117 &        32 & $\ast$ &        1.59 & $\ast$ &        2.59 \\
   10.98 &    \ion{Fe}{23}\,\ion{Ne}{9}\,\ion{Fe}{23}\,\ion{Na}{10}\,\ion{Fe}{17}\, &  7.20 &       102 &        42 & $\ast$ &        1.1 & $\ast$ &        1.67 \\
   12.13 &                      \ion{Ne}{10}\,\ion{Ne}{10}\,\ion{Fe}{17}\,\ion{Fe}{23}\, &  6.75 &       254 &        55 & $\ast$ &       -3.20 & $\ast$ &       -0.23 \\
   12.28 &                                                  \ion{Fe}{21}\,\ion{Fe}{17}\, &  7.05 &        97 &        40 & $\ast$ &        1.06 & $\ast$ &        0.28 \\
   13.45 &                                     \ion{Ne}{9}\,\ion{Fe}{19}\,\ion{Fe}{19}\, &  6.60 &        71 &        40 & $\ast$ &       -2.07 & $\ast$ &       -0.98 \\
   13.52 &                      \ion{Ne}{9}\,\ion{Fe}{19}\,\ion{Fe}{19}\,\ion{Fe}{21}\, &  7.00 &        91 &        37 & $\ast$ &        0.12 & $\ast$ &        0.08 \\
   13.70 &                                                                   \ion{Ne}{9}\, &  6.55 &       161 &        40 & $\ast$ &        1.83 &        &             \\
   14.20 &                                               \ion{Fe}{18}\,\ion{Fe}{18}\, &  6.90 &        93 &        26 & $\ast$ &       -0.48 & $\ast$ &        0.11 \\
   15.01 &                                                                 \ion{Fe}{17}\, &  6.75 &       188 &        30 & $\ast$ &        0.97 & $\ast$ &        0.98 \\
   15.08 &                                                                  \ion{Fe}{19}\, &  6.95 &        64 &        25 & $\ast$ &        2.01 & $\ast$ &        2.06 \\
   15.21 &                                    \ion{Fe}{19}\,\ion{O}{8}\,\ion{O}{8}\, &  6.95 &        63 &        32 & $\ast$ &        1.11 & $\ast$ &        1.61 \\
   15.26 &                                                                 \ion{Fe}{17}\, &  6.75 &        31 &        27 & $\ast$ &       -0.60 & $\ast$ &       -0.50 \\
   16.00 &                                  \ion{Fe}{18}\,\ion{O}{8}\,\ion{O}{8}\, &  6.90 &       103 &        24 & $\ast$ &        1.25 & $\ast$ &        2.97 \\
   16.07 &                                                 \ion{Fe}{18}\,\ion{Fe}{19}\, &  6.90 &        58 &        22 & $\ast$ &        0.65 & $\ast$ &        0.24 \\
   16.78 &                                                                 \ion{Fe}{17}\, &  6.75 &        82 &        24 & $\ast$ &       -0.41 & $\ast$ &       -0.17 \\
   17.05 &                                                 \ion{Fe}{17}\,\ion{Fe}{17}\, &  6.75 &       204 &        33 & $\ast$ &       -0.54 & $\ast$ &        0.23 \\
   18.63 &                                                                   \ion{O}{7}\, &  6.35 &         8 &        12 & $\ast$ &       -0.35 & $\ast$ &        0.59 \\
   18.97 &                                                   \ion{O}{8}\,\ion{O}{8}\, &  6.50 &       373 &        34 & $\ast$ &        3.65 & $\ast$ &        9.77 \\
   21.60 &                                                                   \ion{O}{7}\, &  6.30 &        37 &        32 & $\ast$ &       -1.04 & $\ast$ &        0.92 \\
   21.81 &                                                                   \ion{O}{7}\, &  6.30 &        77 &        35 & $\ast$ &        1.97 &        &             \\
   22.10 &                                                                   \ion{O}{7}\, &  6.30 &        61 &        27 & $\ast$ &        0.60 &        &             \\
   33.73 &                                                       \ion{C}{6}\,\ion{C}{6}\, &  6.15 &        38 &        27 & $\ast$ &        0.03 & $\ast$ &        0.22 \\
 1174.93 &                                                                   \ion{C}{3}\, &  4.95 &         9.6 &         0.8 &        &             & $\ast$ &        0.48 \\
 1175.26 &                                                                   \ion{C}{3}\, &  4.95 &        10.2 &         1.1 &        &             & $\ast$ &        3.23 \\
 1175.71 &                                       \ion{C}{3}\,\ion{C}{3}\,\ion{C}{3}\, &  4.95 &        39.1 &         1.1 & $\ast$ &       -1.55 & $\ast$ &       -1.51 \\
 1176.37 &                                                                   \ion{C}{3}\, &  4.95 &        11.6 &         0.8 &        &             & $\ast$ &        4.00 \\
 1206.50 &                                                                  \ion{Si}{3}\, &  4.80 &        67.3 &         1.1 &        &             &        &             \\
 1218.35 &                                                                     \ion{O}{5}\, &  5.35 &         6.9 &         1.1 & $\ast$ &       -3.12 & $\ast$ &       -0.50 \\
 1238.82 &                                                                     \ion{N}{5}\, &  5.30 &        21.5 &         1.1 & $\ast$ &        0.09 & $\ast$ &       -0.13 \\
 1242.81 &                                                                     \ion{N}{5}\, &  5.30 &        11.8 &         0.9 & $\ast$ &        1.16 & $\ast$ &        1.26 \\
 1253.81 &                                                                    \ion{S}{2}\, &  4.50 &         0.9 &         0.3 & $\ast$ &       -0.02 & $\ast$ &       -0.40 \\
 1264.74 &                                                     \ion{Si}{2}\,\ion{Si}{2}\, &  4.45 &         4.2 &         0.3 & $\ast$ &        2.44 &        &             \\
 1266.11 &                                                                             ...  &  0.00 &         1.1 &         0.2 &        &             &        &             \\
 1294.55 &                                                                  \ion{Si}{3}\, &  4.80 &         1.1 &         0.3 & $\ast$ &        2.68 &        &             \\
 1298.95 &                                                   \ion{Si}{3}\,\ion{Si}{3}\, &  4.80 &         3.0 &         0.6 & $\ast$ &        3.71 &        &             \\
 1309.28 &                                                                   \ion{Si}{2}\, &  4.45 &         1.6 &         0.3 & $\ast$ &       -2.31 &        &             \\
 1323.95 &                                          \ion{C}{2}\,\ion{C}{2}\,\ion{C}{2}\, &  4.75 &         0.8 &         0.3 & $\ast$ &       -0.57 & $\ast$ &        0.71 \\
 1334.53 &                                                                    \ion{C}{2}\, &  4.60 &        32.7 &         0.9 & $\ast$ &      -27.91 & $\ast$ &       -1.84 \\
 1335.71 &                                                       \ion{C}{2}\,\ion{C}{2}\, &  4.60 &        69.5 &         1.2 & $\ast$ &       27.99 & $\ast$ &        1.41 \\
 1351.44 &                                                                             ...  &  0.00 &         3.3 &         0.6 &        &             &        &             \\
 1354.07 &                                                                  \ion{Fe}{21}\, &  7.05 &         5.5 &         0.5 & $\ast$ &        0.88 & $\ast$ &       -0.67 \\
 1364.22 &                                                                             ...  &  0.00 &        -0.4 &         0.3 &        &             &        &             \\
 1371.30 &                                                                     \ion{O}{5}\, &  5.35 &         1.9 &         0.4 & $\ast$ &       -1.56 & $\ast$ &        4.55 \\
 1393.76 &                                                                   \ion{Si}{4}\, &  4.90 &        41.6 &         2.6 & $\ast$ &       -2.54 & $\ast$ &       -1.42 \\
 1401.16 &                                                                    \ion{O}{4}\, &  5.15 &         3.4 &         0.5 & $\ast$ &      -23.68 & $\ast$ &       -0.99 \\
 1402.77 &                                                                   \ion{Si}{4}\, &  4.90 &        23.6 &         0.9 & $\ast$ &       -0.59 & $\ast$ &        1.65 \\
 1407.38 &                                                                    \ion{O}{4}\, &  5.15 &         0.6 &         0.3 &        &             & $\ast$ &       -0.47 \\
 1548.19 &                                                                    \ion{C}{4}\, &  5.05 &        83.0 &         2.4 & $\ast$ &        0.87 & $\ast$ &       -0.73 \\
 1550.78 &                                                                    \ion{C}{4}\, &  5.05 &        44.8 &         2.0 & $\ast$ &        0.91 & $\ast$ &        1.57 \\
\hline
\end{tabular}
\end{center}
$^a$~Wavelengths (\AA).
$^b$~Multiple identifications indicate unresolved lines in the CHIANTI database which contribute to the measured flux.
$^c$~Temperature (K) of maximum emissivity.
$^d$~Measured fluxes (${\rm 10^{-16}\,erg\,s^{-1}\,cm^{-2}}$) with uncertainties at the 68\% confidence level.
$^e$~Lines selected for the EMD reconstruction with each method.
$^f$~Comparison between observed and predicted line fluxes (method 1).
$^g$~Comparison between observed and predicted line fluxes (method 2).
\normalsize
\end{table*}

\subsection{HARPS-N observations} \label{sec:harpsn}
We observed \target\ within a dedicated DDT program (ID: A42DDT5, PI: A. Maggio) at the Telescopio Nazionale Galileo (TNG, La Palma, Canary Islands) by using the HARPS-N high-resolution spectrograph \citep{Cosentino2014} in the visible band (383 - 690 nm). We obtained six spectra of the target between Feb 23rd and Feb 26th, 2021, in order to perform a follow-up of the first \xmm\ observation through the monitoring of the chromospheric emission in the \ion{Ca}{2} H\&K lines. Three additional spectra of \target\ were collected between Aug 23rd and Aug 26th to support the coordinated \xmm-HST observation within the Global Architecture of Planetary Systems (GAPS) project. Since \target\ was also included in the \textit{Young Objects} sample of GAPS \citep{Carleo2020}, a large amount of data (almost 190 spectra between Mar 2019 and Mar 2022) has been collected by this observing program, aiming to measure the masses of the four planets of the system. For all datasets, we calculated the values of the $\log R^{\prime}_{\rm HK}$ activity index by using the procedure described in \citet[and references therein]{2011arXiv1107.5325L} available on the YABI workflow interface, implemented at the INAF Trieste Observatory\footnote{\url{https://www.ia2.inaf.it/}}.

\begin{figure}
    \centering
    \resizebox{0.46\textwidth}{!}{\includegraphics[trim={3.0cm 2.9cm 2.5cm 2.9cm},angle=270]{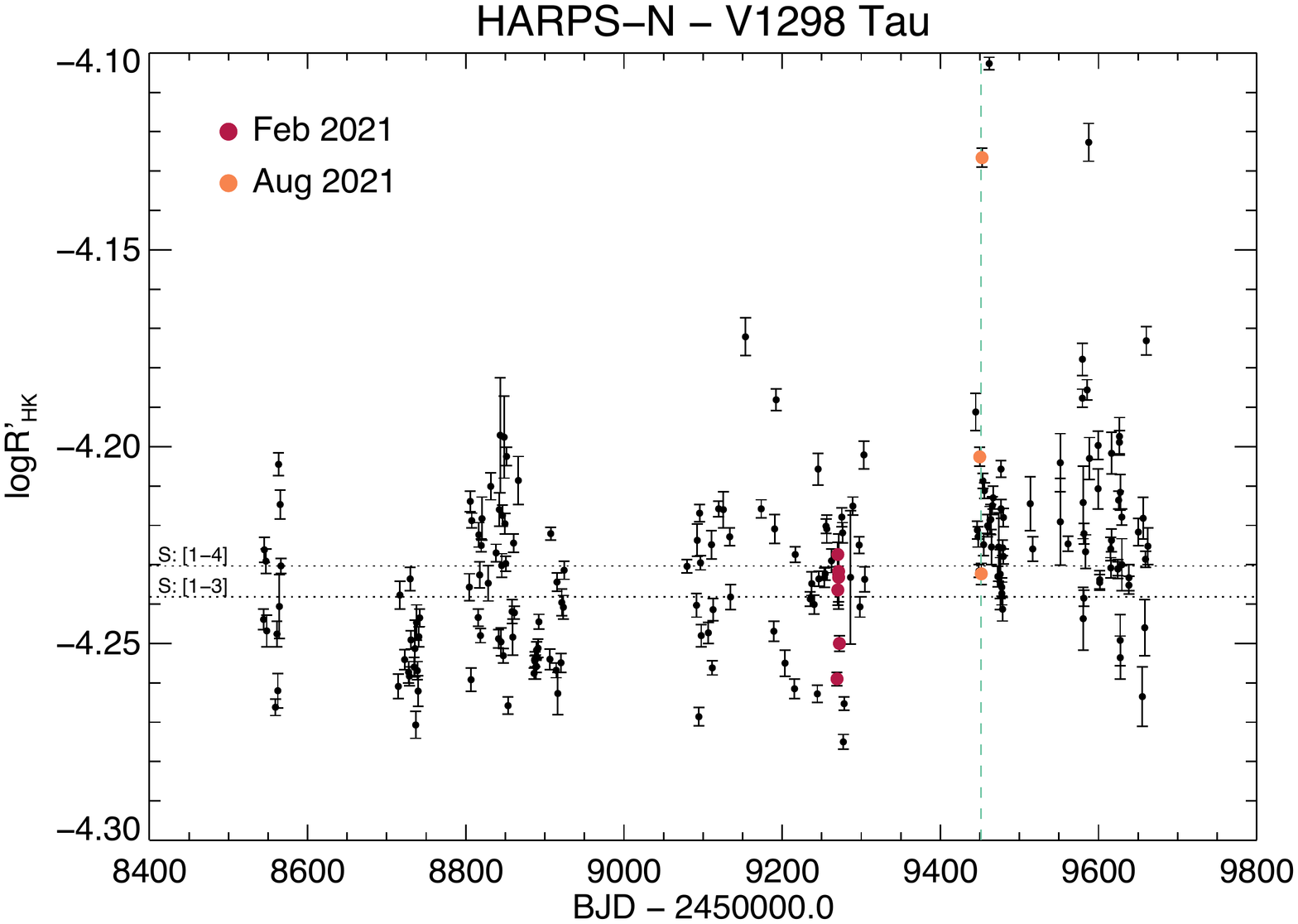}}
    \caption{Time series of the $\log R^{\prime}_{\rm HK}$ activity index obtained with HARPS-N@TNG within the framework of the GAPS observing program (black dots, $1 \sigma$ errors), and within the dedicated follow-up programs of the XUV observations (see the legend). The dotted horizontal lines represent the median level in the four observing seasons (labelled as {\tt S:[1-4]}, i.e.\ the complete time series) compared to the first three ({\tt S:[1-3]}. The blue vertical dashed line indicates the date of the simultaneous \xmm-HST observations.}
    \label{fig:hk}
\end{figure}

The use of the GAPS data allows us to perform a comparison of the values of $\log R^{\prime}_{\rm HK}$ obtained during the \xmm-HST observations with its typical behaviour over four years. Fig.\ref{fig:hk} shows the time series of this chromospheric activity index obtained as part of the GAPS program (black dots), and the values taken during the DDT program in Feb 2021 (red dots) and the additional observations in Aug 2021 (orange dots). No significant difference is visible between our first set of DDT data and the GAPS ones, since the chromospheric indexes appear well within the mean distribution. Instead, we observe an excess in the value of $\log R^{\prime}_{\rm HK}$ ($> -4.15$) for one spectrum of the second set of DDT data (on 2021 Aug 26 at 05:16 UT) and for two GAPS spectra (on 2021 Sep 4 at 04:51 UT, and on 2022 Jan 08 at 01:03 UT, respectively). 
We note that the mean value of the $\log R^{\prime}_{\rm HK}$ appears slightly larger in the last season than in the previous ones, as indicated by the median values drawn in Fig. \ref{fig:hk} as horizontal lines, suggesting an enhancement of the stellar activity, in full agreement with what we found in X-rays. We hypothesize that the spectra showing the highest values of $\log R^{\prime}_{\rm HK}$ could have caught the star during large stellar flares. For a double check, we extracted the H$\alpha$ index from those spectra by using the ACTIN code \citep{gomesdasilva18}. Actually, we observed a higher value also for this chromospheric proxy, which corroborates the flare hypothesis (see e.g. \citealt{dimaio2020}), although the optical spectra show no significant emission in the H$\alpha$ line core.

Finally, we used a subset of these HARPS-N data to obtain a co-added spectrum aiming to perform a comparison between the photospheric abundances with those in the outer stellar atmosphere probed by the XUV lines. To derive an estimate of the metallicity we performed the comparison with synthetic spectra (calculated using MOOG and a grid of Kurucz  model atmospheres), which include \ion{Fe}{1} lines. We obtained T$_{\rm eff} = 4980$\,K, $\log g = 3.95 \pm 0.15$ dex, $V_{\rm microt} = 0.85$\,km/s, and the resulting [Fe/H] is $0.00 \pm 0.15$ dex (with $\log \epsilon(Fe)_\odot = 7.50)$, in agreement with the solar abundance reported by \cite{natpaper}.
The combination of relatively high rotational velocity and low effective temperature prevents us from
deriving accurate abundances of other elements, because of possible unknown line blending and NLTE effects for carbon, silicon, and magnesium (see also \citealt{natpaper}).
 
\section{Results}
\subsection{Source variability}
\label{sect:lc}
The EPIC X-ray light curves of \target\ (Fig. \ref{fig:xraylc}) show low-level but significant time variability in the first 15\,ks of the Aug 2021 XMM-Newton observation, with a peak count rate $\sim 60$\% higher with respect to the average quiescent level observed in the next 21\,ks time segment. This behavior is more clear in the MOS light curves, less affected by high background contamination than pn. 
The simultaneous NUV light curve obtained with the OM shows just some flickering at the level of $\sim 30$\% with respect to the average count rate. 

Although no large flare is evident, we analyzed separately the EPIC spectra accumulated in the two time segments, which we dubbed "quiescent" and "high-state", in order to check for possible variations of the characteristics of the coronal plasma. Further assessment of the long-term variability of \target\ is postponed to the end of Sect.\ref{sect:xspec}.

\begin{splitdeluxetable*}{cccccccccccBcccccccccc}
\tablecaption{Best fit parameters from modelling of the EPIC and RGS spectra of \target. 
\label{tabxspec}}
\tabletypesize{\scriptsize}
\tablewidth{0pt}
\tablehead{
\colhead{Obs Id} & \colhead{T1} & \colhead{EM1} & \colhead{T2} & \colhead{EM2} & \colhead{T3} & \colhead{EM3} & \colhead{$\chi^2$} & \colhead{d.o.f.} &
\colhead{$f_\mathrm X$} & 
\colhead{$L_\mathrm X$} & & \multicolumn{9}{c}{Abundances and 90\% confidence ranges (solar units; Anders \& Grevesse 1989)} \\
 & \colhead{$10^6$\,K} & \colhead{$10^{52}$\,cm$^{-3}$} & \colhead{$10^6$\,K} & \colhead{$10^{52}$\,cm$^{-3}$} & \colhead{$10^6$\,K} & \colhead{$10^{52}$\,cm$^{-3}$} & & & 
\colhead{$10^{-12}$\,erg s$^{-1}$ cm$^{-2}$} & \colhead{$10^{30}$\,erg$~$s$^{-1}$} & &
Mg               & Fe                  & Si       & S           & C                 & O            & N          & Ca         & Ne \\
& & & & & & & & & & & FIP (eV) &
7.65 & 7.90 & 8.15 & 10.36 & 11.26 & 13.62 & 14.53 & 15.76 & 21.56
}
\startdata
\\
Feb 2021 & $2.9^{+0.6}_{-0.3}$ & $2.7^{+0.7}_{-0.7}$      & $8.0^{+0.3}_{-0.5}$ & $6.7^{+1.9}_{-1.5}$            & $16.5^{+2.7}_{-1.9}$ & $3.9^{+0.9}_{-1.0}$         & 532.8   & 431  & $1.24^{+0.01}_{-0.02}$ & $1.74^{+0.01}_{-0.03}$ &
Feb 2021  & 0.33 & 0.17 & 0.25 & = O & =O & 0.29 & =O & = Ne & 0.95 \\
& & & & & & & & & & & & $[0.20,0.50]$ & $[0.13,0.22]$ & $[0.15,0.38]$ & & & $[0.23,0.37]$ & & & $[0.95,1.63]$ \\
Quiescent & $4.9^{+0.8}_{-0.6}$ & $5.8^{+1.5}_{-1.6}$     & $10.8^{+0.6}_{-0.7}$ & $13.1^{+1.1}_{-3.0}$          & $43.5^{\rm unbound}_{-23.4}$ & $1.3^{+2.2}_{-0.8}$             & 605.0   & 510  & $1.52^{+0.01}_{-0.08}$ & $2.13^{+0.02}_{-0.11}$ &
Quiescent & 0.19 & 0.10 & 0.10 & = O & =O & 0.22 & =O & = Ne & 0.47 \\
& & & & & & & & & & & & $[0.09,0.28]$ & $[0.08,0.12]$ & $[0.03,0.17]$ & & & $[0.16,0.29]$ & & & $[0.37,0.70]$ \\
High-state & $4.3^{+0.8}_{-0.7}$ & $4.7^{+1.4}_{-1.3}$     & $10.7^{+0.7}_{-0.7}$ & $8.7^{+3.6}_{-2.5}$          & $24.3^{+6.5}_{-3.1}$ & $8.8^{+1.7}_{-2.5}$         & 474.8   & 422  & $2.04^{+0.01}_{-0.05}$ & $2.85^{+0.02}_{-0.07}$ &
High-state & 0.25 & 0.19 & 0.08 & = O & =O & 0.38 & =O & = Ne & 0.65 \\
& & & & & & & & & & & & $[0.07,0.47]$ & $[0.14,0.26]$ & $\le 0.23$ & & & $[0.28,0.51]$ & & & $[0.39,0.97]$ \\
Combined  & $4.8^{+0.5}_{-0.5}$ & $5.3^{+0.8}_{-0.8}$     & $10.3^{+0.5}_{-0.5}$ & $9.0^{+1.8}_{-1.5}$          & $24.3^{+13.9}_{-4.1}$ & $3.4^{+1.2}_{-1.4}$         & 1645.6 & 1315 & $1.46^{+0.01}_{-0.02}$ & $2.05^{+0.01}_{-0.03}$ &
Combined  & 0.21 & 0.13 & 0.15 & 0.26 & =O & 0.28 & =O & = Ne & 0.61 \\
& & & & & & & & & & & & $[0.14,0.29]$ & $[0.11,0.15]$ & $[0.10,0.21]$ & $[0.13,0.40]$ & & $[0.24,0.33]$ & & & $[0.50,0.73]$ \\
& & & & & & & & & & & EMD method 1 & 0.32 & 0.13 & 0.81 & 0.23 & 0.18 & 0.20 & 0.41 & & 0.91 \\
& & & & & & & & & & & & $[0.09,1.20]$ & $[0.09,0.19]$ & $[0.51,1.29]$ & $[0.12,0.45]$ & $[0.13,0.26]$ & $[0.12,0.35]$ & $[0.35,0.48]$ & & $[0.47,1.78]$ \\
& & & & & & & & & & & EMD method 2 & 0.41 & 0.15 & 1.49 & 0.35 & 0.25 & 0.03 & 0.23 & & 0.59 \\
& & & & & & & & & & & & $[0.10,0.55]$ & $[0.14,0.16]$ & $[1.45,1.50]$ & $[0.24,0.41]$ & $[0.24,0.26]$ & $[0.02,0.04]$ & $[0.20,0.28]$ & & $[0.46,1.68]$ \\
\enddata
\tablecomments{Unabsorbed X-ray flux and luminosity in the 0.1--10 keV band. Errors are quoted at the 90\% confidence level.}
\end{splitdeluxetable*}

\pagebreak
\subsection{Global fitting of X-ray spectra}
\label{sect:xspec}
For the spectral analysis, performed with {\sc xspec} V12.12.0, we proceeded as in \citet{Maggio+2022}. Initially we applied a best-fitting procedure only to the EPIC (MOS1, MOS2, and pn) X-ray spectra. We adopted an optically-thin coronal emission model (AtomDB v3.0.9, \citealt{aped}) composed of three isothermal components (3T, {\sc vapec}), with the abundances of all elements linked to the iron abundance. Next, we added also the combined RGS1 and RGS2 high-resolution spectra, and allowed up to nine elements as free parameters: C, N, O, Ne, Mg, Si, S, Ar, and Fe.
A global interstellar absorption was also included in the model with a multiplicative component ({\sc phabs}).

Eventually, we reduced the number of free parameters by fixing the interstellar hydrogen column density, $N_{\rm H}$, and the abundances of C, N and Ar, because these parameters were poorly constrained. In fact, the best-fit value of $N_{\rm H}$ was consistent with the value derived from the known B-V color excess, $E(B-V) = 0.024 \pm 0.015$ (\citealt{David+2019b}),
which implies an extinction $A_{\rm V} = 0.074 \pm 0.05$ and $N_{\rm H} = 1.6\pm1.1 \times 10^{20}$\,cm$^{-2}$, hence we fixed this parameter to the nominal value. Similarly, the best-fitting procedure yields just uninformative upper limits for both the C and N abundances, and we decided to fix both of them to the Oxygen abundance, guided by the similarity of the First Ionization Potentials of these elements (see \ref{sect:abund}). The line complex of \ion{Ar}{17} and \ion{Ar}{18} at 3.14\,keV and 3.32\,keV, respectively, is quite evident in the combined spectrum (Fig. \ref{fig:xraylc}), and the best-fit abundance is near the solar value, but we fixed it to the abundance of Neon due to the large uncertainties.

\begin{figure}
    \centering
    \resizebox{0.46\textwidth}{!}{\includegraphics[trim=1.2cm 13.5cm 3.5cm 4cm, clip]{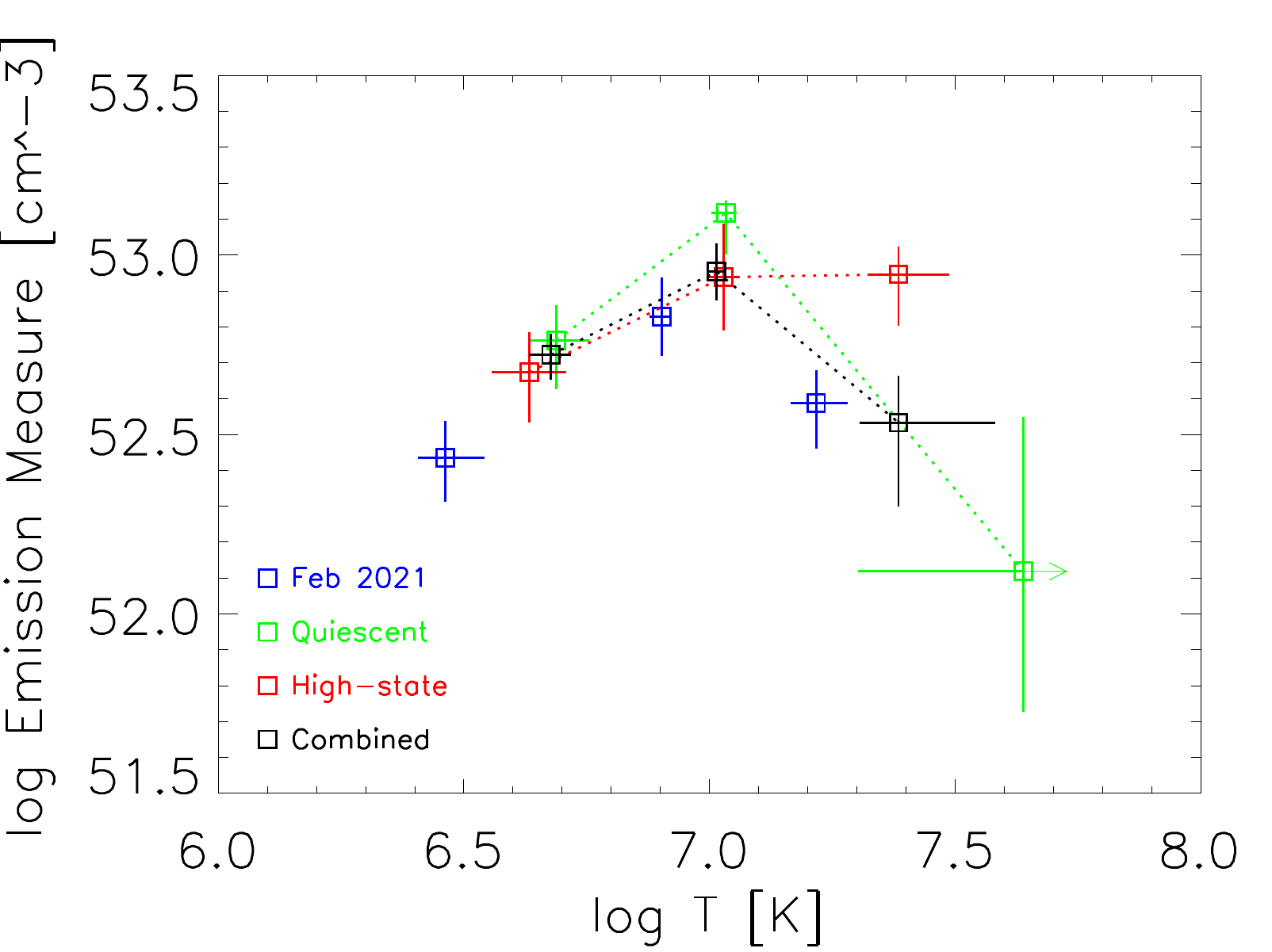}}
    \caption{Volume emission measure vs.\ plasma temperature resulting from the spectral analysis of the XMM-Newton spectra with 3T models. The results from both the Feb 2021 and Aug 2021 observations are shown, the latter splitted into a "quiescent" and a "high-state" phase. 
    }
    \label{fig:multit}
\end{figure}

Table \ref{tabxspec} reports the parameters from the best-fit models to EPIC and RGS
spectra of the "quiescent" and "high-state" time segments. We also show the results of the fitting of the Feb 2021 spectra and of the combined spectra obtained by adding the Feb 2021 and Aug 2021 data, performed with the same procedure. During the Aug 2021 observation, the coronal plasma of \target\ was characterized by a cold temperature component at 4--5\,MK, and a second component at 10\,MK. The volume emission measures of these components did not change appreciably from the "quiescent" to the "high-state" phase, and they appear similar also to those probed by the Feb 2021 observation. Moreover, there is evidence in all cases of very hot plasma, at 15--25\,MK, which is responsible for most of the observed variability: this component has a very low emission measure during the "quiescent" phase in Aug 2021, while it becomes the dominant one during the "high-state" phase.
A comprehensive view of the results of these multi-component isothermal fitting of the XMM spectra is displayed in Fig.\ref{fig:multit}.
This is typical behavior for the variability of the coronal plasma in young active stars.

The unabsorbed flux and the luminosity of \target\ in the quiescent phase are $f_{\rm x,q} = 1.4^{+0.1}_{-0.2}\times10^{-12}$\,\fxu\ and $L_{\rm x,q} = 2.0^{+0.2}_{-0.3}\times10^{30}$\,\lxu, respectively in the band 0.1--2.4\,keV. These values are about 20\% higher than those observed in Feb 2021. A further increase of $\ga 30$\% occurred during the variable phase in Aug 2021, characterized by an average flux and luminosity of $f_{\rm x,v} = 1.82^{+0.03}_{-0.08}\times10^{-12}$\,\fxu\ and $L_{\rm x,v} = 2.56^{+0.04}_{-0.11}\times10^{30}$\,\lxu. A slightly larger variability can be derived in the broader X-ray band 0.1--10\,keV (Table \ref{tabxspec}). Considering also previous ROSAT and Chandra observations \citep{Poppenhaeger+2020}, the X-ray emission of \target\ shows a long-term variability within a factor $\sim 2$.

Finally, we evaluated that the X-ray to bolometric luminosity ratio, $\log L_{\rm x}/L_{\rm bol}$, ranged from -3.35 to -3.15 in the time span between Feb 2021 and Aug 2021, and the flux at the stellar surface, $F_{\rm x}$, was between $1.7 \times 10^7$ and $2.6 \times 10^7$\,\fxu. These surface fluxes are 10--100 times higher than in the case of the Sun at the maximum and minimum of its magnetic activity cycle \citep{Chadney+2015}.

The significant but low-amplitude variability of \target\ is also confirmed by considering the 3-year time series of the chromospheric \ion{Ca}{2} H\&K emission lines (Fig.\ref{fig:hk}), where the $\log R^{\prime}_{\rm HK}$ index shows a full range of variation $\la 0.2$\,dex. These measurements can be converted into the pure activity-related R$^{+}_{\rm HK}$ index by \citet{Mittag+2013}, and we have employed the latter to predict the X-ray to bolometric luminosity ratios by means of the chromospheric to coronal flux-flux relationship derived by \citet{Fuhrmeister+2022} for active G-type stars. We found a full range of $\log L_{\rm x}/L_{\rm bol}$ between $-3.7$ and $-3.2$, i.e. a possible variability of about a factor 3.

\begin{figure}
    \centering
    \resizebox{0.46\textwidth}{!}
    {\includegraphics[trim=1.8cm 13.5cm 4.0cm 4cm, clip]
    {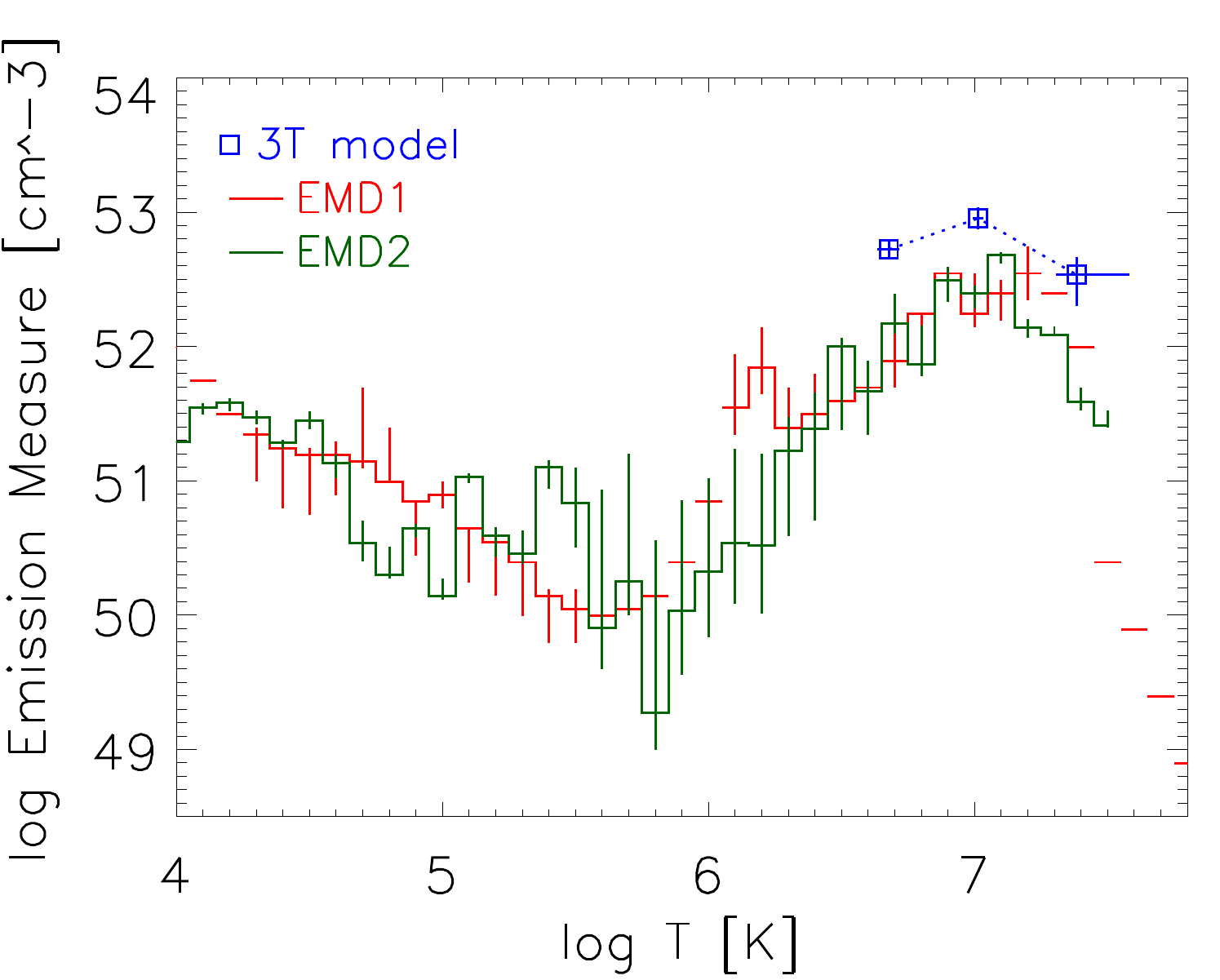}}
    \resizebox{0.46\textwidth}{!}
    {\includegraphics{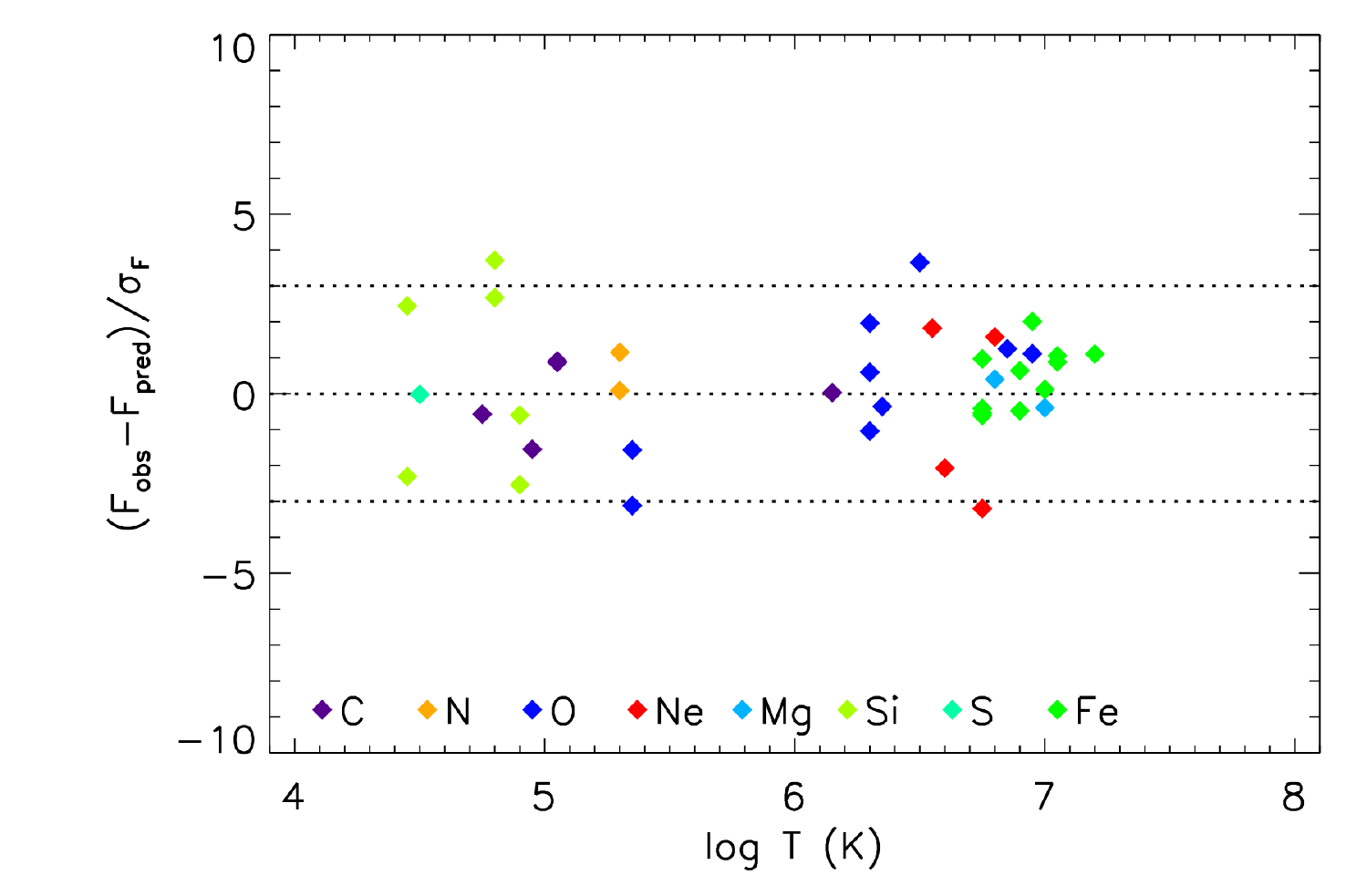}}
    \resizebox{0.46\textwidth}{!}
    {\includegraphics{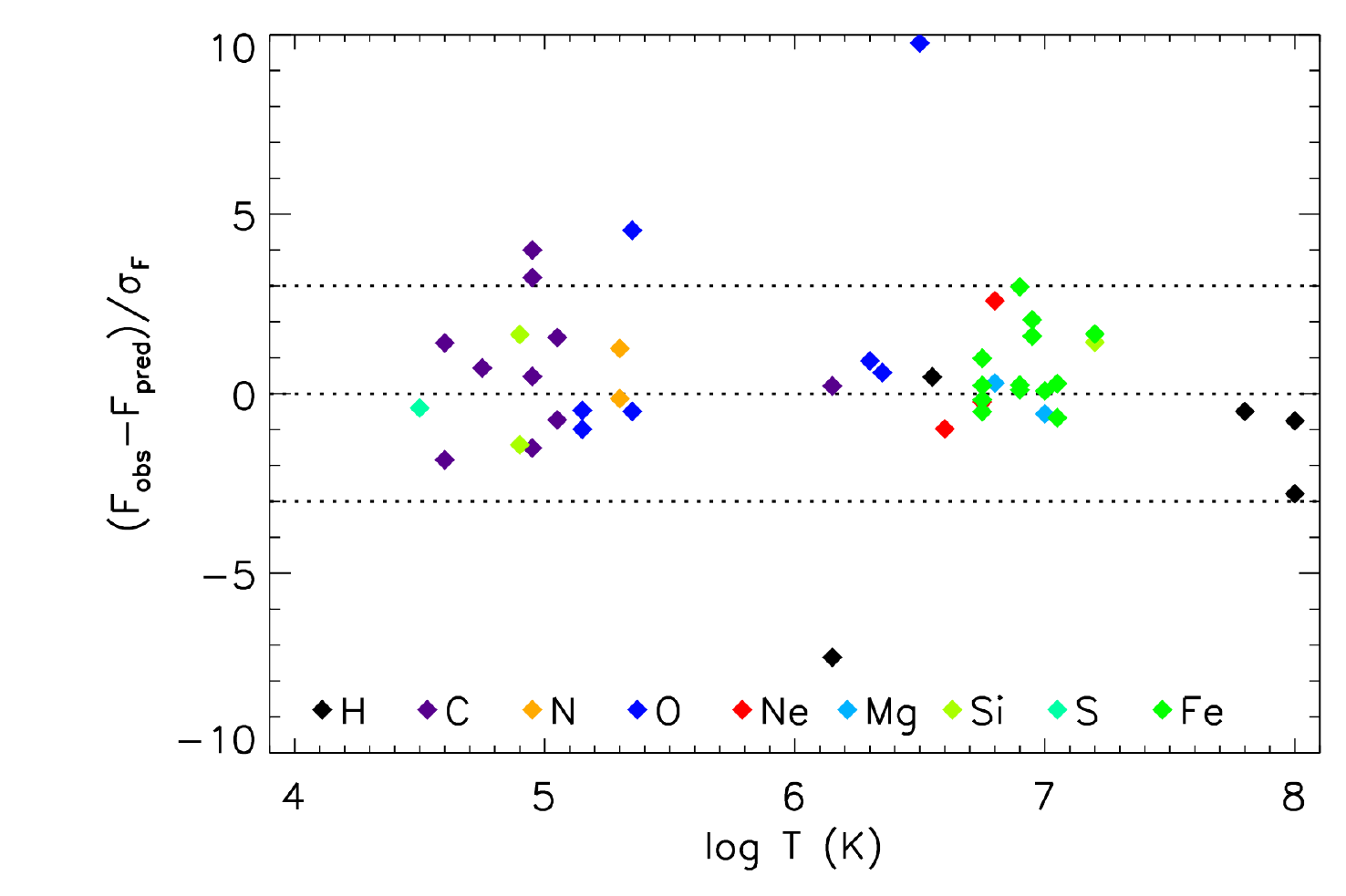}}
    \caption{Top: Plasma Emission Measure Distributions vs.\ temperature resulting from the joint analysis of XMM-Newton/RGS and HST/COS spectra, with method 1 (red) and method 2 (green), compared with the 3-T model best-fitting the combined EPIC and RGS spectra. The 3-T EM values appear higher because each of them corresponds to several temperature bins in the EMDs. Middle and bottom: Differences, in $\sigma$ units, between measured line fluxes and values predicted with the two methods, vs. temperature at the peak emissivity. In the bottom panel (method 2), the black "H" symbols represent narrow-band measurements of the X-ray continuum.}
    \label{fig:emds}
\end{figure}

\subsection{Emission Measure Distribution}\label{sect:dem}

The full set of UV and X-ray line fluxes probes emission from material with temperatures ranging from $3\times10^4$\,K to $2\times10^7$\,K. This line set allows us to derive the Emission Measure Distribution (EMD) vs.\ temperature of the entire stellar atmosphere, from the chromosphere to the corona. To this aim, we assume that the emission is due to a collisionally-excited optically-thin plasma. We have checked that deviations from this hypothesis affect only marginally some strong lines which form at chromospheric temperatures (see Appendix \ref{Appendix}).

The lack of significant variability of the chromospheric $R^{\prime}_{\rm HK}$ index and of the coronal emission between Feb 2021 and Aug 2021 observations (Sect.\ \ref{sect:lc}), except for the thermal components hotter than $\sim 10$\,MK, supports our assumption that the same occurred for the strength of the EUV emission lines.

Interstellar absorption, assuming $N_{\rm H} = 1.6\times 10^{20}$\,cm$^{-2}$, was properly taken into account for computing unabsorbed X-ray line fluxes, while we have employed the \citet{Fitzpatrick1999} extinction law for the UV lines.

We employed two different and independent methodologies for the reconstruction of the EMD vs.\ temperature. 
We have first derived the EMD following \citet{san03} (Method 1). An initial EMD with 0.1 dex resolution in temperature, based on global fitting results, is used to calculate the expected line fluxes for the source. These calculations include contributions from all lines in identified blends, according to the AtomDB (v3.0.9) database. The ratio between observed and predicted line fluxes is used to modify the EMD and the abundances of the elements in an iterative process, thus obtaining the best result to minimize these line ratios.
The EMD is computed initially using abundances relative to iron. At the end of the process, the whole EMD is shifted by +0.89 dex, to correct for the $[Fe/H] = -0.89$ value determined with the global fitting of the combined spectrum (Sect.\ \ref{sect:xspec}).
The method provides also uncertainties on both EMD and abundances with a MonteCarlo method which takes into account the uncertainties on the measured line fluxes, but keeping fixed the iron abundance. 

In this process, special attention was made to the consistency of spectral line fluxes of similar ions or temperature of formation, excluding some of them (e.g.\ \ion{Si}{3}~1206.50\,\AA). A special treatment was also applied to the \ion{C}{3} multiplet at $\sim 1176$\,\AA, having inaccurate atomic data in AtomDB v3.0.9. In this case, we adopted the \citet{raym88} atomic data to evaluate the flux of the whole multiplet.

Then, we derived an alternative solution for the plasma EMD and abundances by employing the MCMC approach implemented in the PINTofALE software suite (v2.954)  \citep{Kashyap1998}. 
In applying this procedure (Method 2) we  adopted the CHIANTI (v7.13) atomic database, since it appears more reliable than APEC in reproducing some strong lines in the UV range. We considered a subset of measured line fluxes, as reported in Table~\ref{tab:linefluxes}. In particular, we did not consider density-sensitive lines, lines with uncertain identification, and lines whose fluxes appear incompatible among themselves in the hypothesis of collisionally-excited plasma\footnote{This issue occurred for some Si lines in the UV band, possibly because the population of the upper levels from which they originate have significant contributions also from other mechanisms, like recombination.}.
To obtain absolute abundances and to better constrain the hottest components of the EMD we complemented these selected line fluxes with the measurements of the total flux (lines + continuum) in five wavelength intervals, including also intervals at short wavelengths where the emission is dominated by the hot plasma (\citealt{Pillitteri+2022}). To this aim we selected the intervals reported in Table~\ref{tab:bandfluxes}. The observed fluxes in these intervals were obtained from the EPIC data with XSPEC. We computed the total emissivities associated to these measurements including, in addition to the continuum contribution, the contribution of all the emission lines contained in the considered wavelength interval. We run the MCMC procedure several times, adjusting at each step this emissivity on the basis of the inferred abundances. 

Considering the temperature ranges covered by the emissivities of the selected lines, we derived the EMDs over a temperature grid ranging from $\log T = 4.0$ to $\log T = 7.5$, with $\Delta \log T = 0.1$. 
The results are shown in Fig.\ref{fig:emds}, together with plots of the residuals between observed and predicted line fluxes with the two methods (Tab.\ \ref{tab:linefluxes}).

The EMD obtained with method 1 appears smoother than the other, but the two solutions turned out to be consistent within $1 \sigma$ uncertainties, except for a few bins, in spite of a slightly different set of emission lines and methodologies of EMD reconstruction. Note that method 1 joins the estimated emission measure values between $\log T = 5.5$ and 6.1\,K with a few bins without formal errors, because this is the temperature range less constrained by the available UV and X-ray emission lines. The same occurs for the EMD at $\log T \ge 7.3$\,K. Method 2 extends the reconstructed EMD up to $\log T = 7.5$\,K thanks to the inclusion of the narrow band X-ray fluxes which inform on the continuum emission level. 
Moreover, the 0.1--2.4\,keV broad band flux and the Li-like lines of C and N have emissivity functions with significant contributions also from plasma in the $\log T$ range 5.5--5.9\,K. It is because of these emissivities that the EMD can be constrained with Method 2 also in this temperature range, although with large error bars.

In Appendix \ref{appendix2} we show a comparison of the EMD of \target\ with those of other G--K stars having different activity levels.

\begin{table}
\label{tab:bandfluxes}
\caption{Measured X-ray fluxes in selected wavelength bands of V1298~Tau.}
\scriptsize
\begin{center}
\begin{tabular}{r@{--}lr@{--}lcr@{$\;\pm\;$}lc}
\hline\hline
\multicolumn{2}{c}{$\lambda^a$}      & \multicolumn{2}{c}{$E^b$} & $\log T^c_{\mathrm max}$ & \multicolumn{2}{c}{$F^d$} & $\frac{(F_{obs}-F_{pred})^e}{\sigma}$\\ 
\hline
 2.48 &   4.13  & 3.00 & 5.00 & 8.00 &   46.8 & 2.9 & -0.76 \\
 4.13 &   5.17  & 2.40 & 3.00 & 8.00 &   39.2 & 1.9 & -2.79 \\
 8.49 &   8.92  & 1.41 & 1.46 & 7.80 &   14.9 & 0.3 & -0.49 \\
27.55 &  30.24  & 0.41 & 0.45 & 6.55 &   28.6 & 0.5 &  0.47 \\
 5.17 & 123.99  & 0.10 & 2.40 & 6.15 & 1040.0 & 10  & -7.34 \\
\hline\\
\end{tabular}
\end{center}
\scriptsize
$^a$~Wavelength range (\AA).
$^b$~Energy range (keV).
$^c$~Temperature (K) of maximum emissivity.
$^d$~Observed fluxes (${\rm 10^{-15}\,erg\,s^{-1}\,cm^{-2}}$).
$^e$~Comparison between observed and predicted line fluxes. \\
\normalsize
\end{table}

\begin{figure}
    \centering
    \resizebox{0.46\textwidth}{!}
    {\includegraphics[trim=1.5cm 13.5cm 4.0cm 4cm, clip]
    {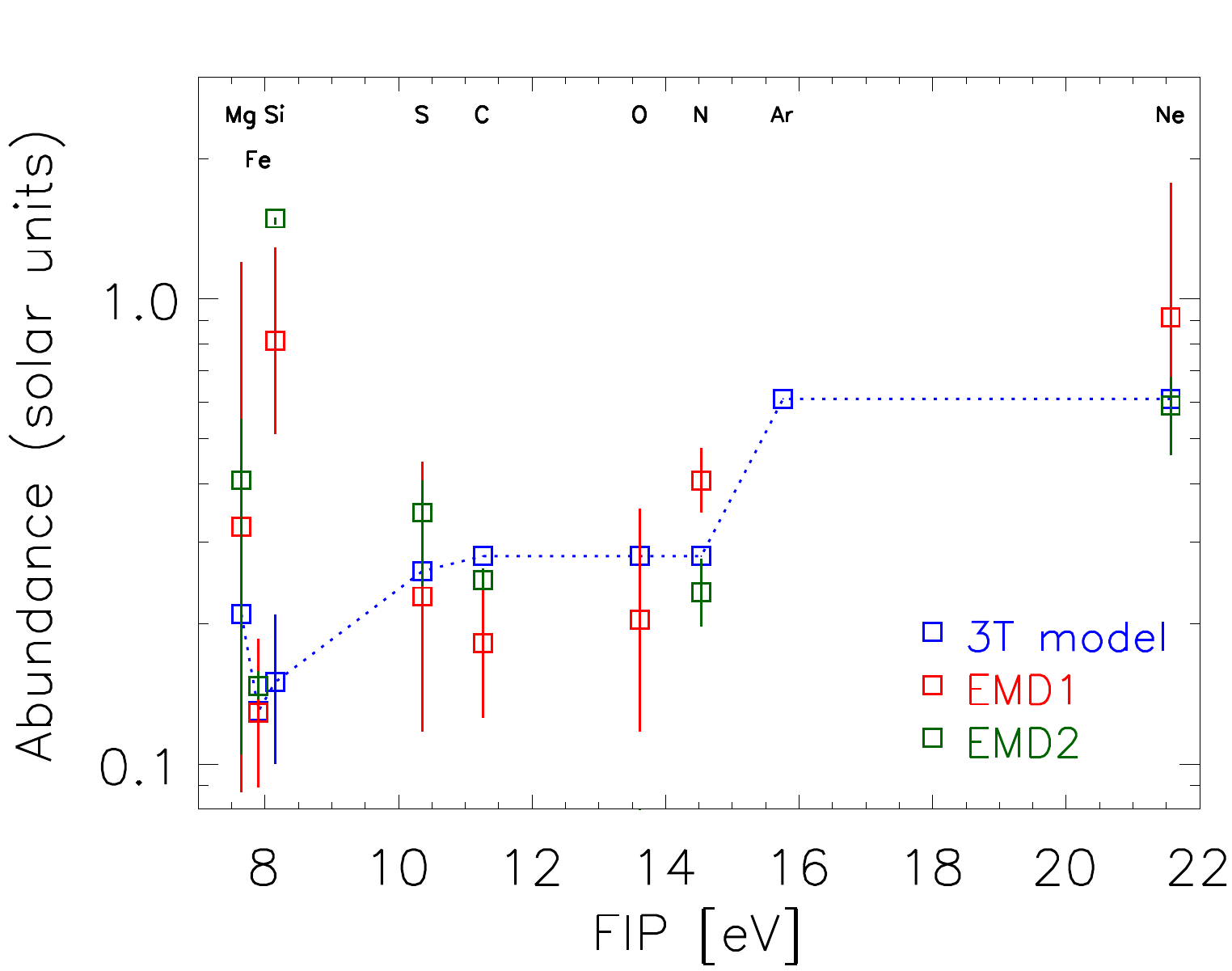}}
    \caption{Chemical abundances vs.\ First Ionization Potential for the global fitting of the combined X-ray spectrum, and the EMD reconstructions with method1 and method 2.}
    \label{fig:abund}
\end{figure}

\begin{figure*}
    \centering
    \resizebox{0.495\textwidth}{!}{\includegraphics[trim=3.2cm 14.0cm 2.3cm 3.5cm, clip]{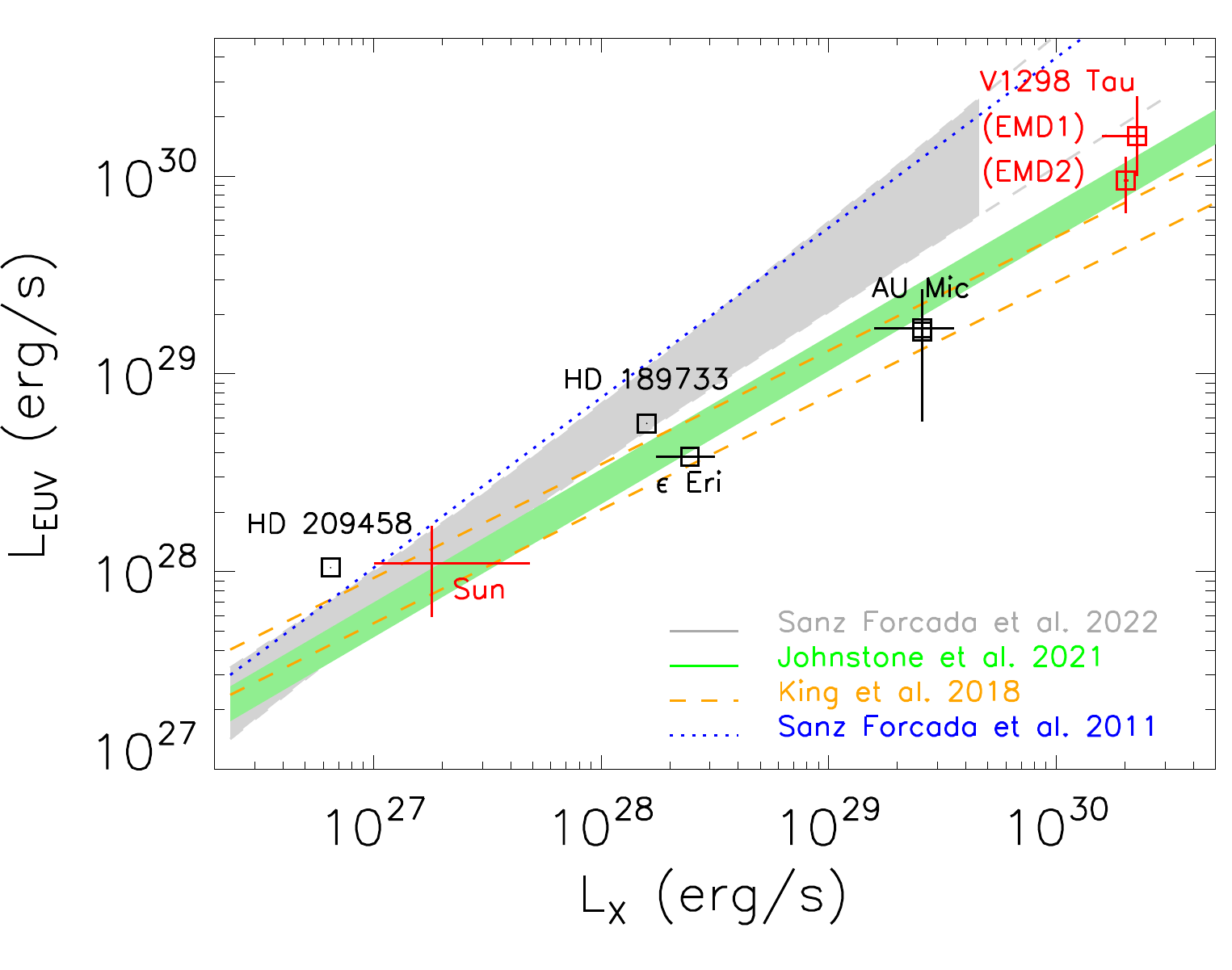}}
    \resizebox{0.495\textwidth}{!}{\includegraphics[trim=3.1cm 14.0cm 2.3cm 3.5cm, clip]{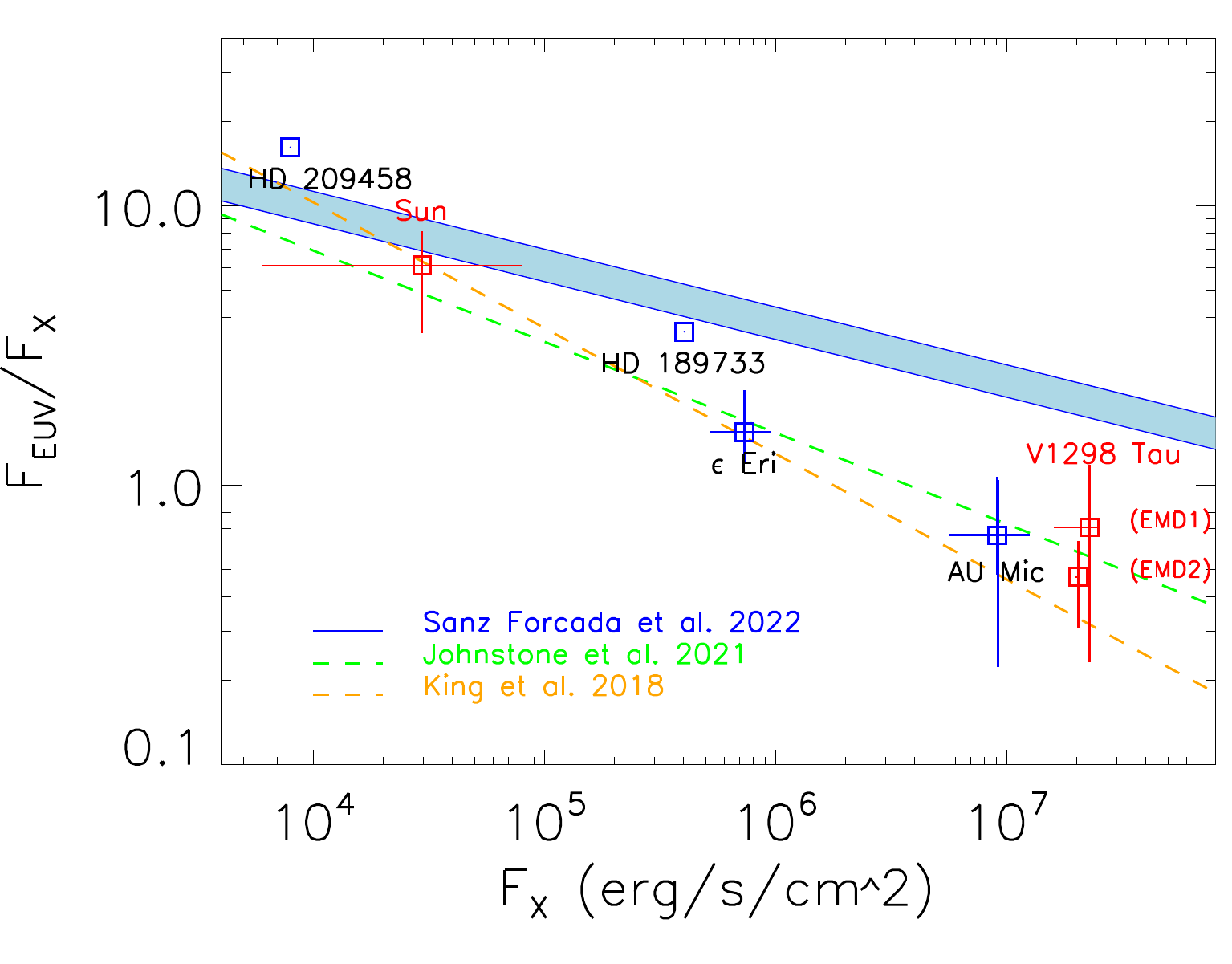}}
    \caption{(Left) Comparison of different X-ray (5--100\,\AA) to EUV (100--920\,\AA) luminosity scaling laws (see labels) with measurements for \target, the Sun, and other benchmark stars with exoplanets. The gray band indicates the 95\% confidence region relative to the SF22 scaling law, and the gray dashed lines its extrapolation. The green band spans over stars with radii in the range 0.7--1.3\,$R_{\odot}$, according to J21. The blue dotted line represents the old SF11 scaling law, while the orange dashed lines refer to K18. (Right) Analogous plot for the scaling law of the EUV to X-ray ratio vs.\ X-ray flux at the stellar surface. The old SF11 law was omitted for clarity, while the new SF22 law is computed assuming the same two stellar radii as above.
    }
    \label{fig:xuvscale}
\end{figure*}

\subsection{Chromospheric and coronal abundances}
\label{sect:abund}
The elemental abundances in the corona of \target, derived from the global analysis of the combined XMM-Newton spectra (Sect. \ref{sect:xspec}), are shown in Fig.\ref{fig:abund} as a function of the First Ionization Potential (FIP). Similar values were obtained also for the single observations in Feb 2021 and Aug 2021, as well as for the quiescent and high-state segments (Tab.\ \ref{tabxspec}). Elements with low FIP (Fe, Mg, and Si) are systematically underabundant with respect to elements with $FIP \ga 12$\,eV, such as O, N, or Ne. This trend, dubbed inverse FIP effect, is typical of young active stars \citep{Maggio+2007,Scelsi+2007}.

In the same figure are shown the values derived together with the EMDs, which resulted in fair agreement with the values from the global fitting of X-ray spectra, except for the case of Silicon. In this case, we are assuming that the abundances of elements such as O and Si remain constant in the full range of temperatures explored, but we recognize that the abundances could change somewhere between the chromosphere and the corona.
Moreover, some differences may be due to the two different atomic databases employed for the analysis of the emission lines.

We recall also that the abundances of \target\ in the chromosphere and corona are scaled with respect to the solar photospheric abundances \citep{AG1989}. Stellar abundances for each element should be employed for a proper assessment of any trend with the FIP \citep{SanzForcada+2004}, but unfortunately, accurate photospheric abundances cannot be determined for \target (Sect.\ref{sec:harpsn}), except for the iron. In this latter case, the difference between the photosphere and the corona is clearly established.


\pagebreak
\subsection{EUV vs.\ X-ray scaling laws}

We have employed the EMDs presented above to compute the X-ray luminosity in the 5--100\,\AA\ band, and the EUV luminosity in the 100--920\,\AA\ band. We have obtained $L_{\rm X} = 2.26 \times 10^{30}$\,erg s$^{-1}$ and $L_{\rm EUV} = 1.60^{+0.95}_{-0.60} \times 10^{30}$\,erg s$^{-1}$ with method 1, and $L_{\rm X} = (2.02 \pm 0.05) \times 10^{30}$\,erg s$^{-1}$ and $L_{\rm EUV} = (0.95 \pm 0.30) \times 10^{30}$\,erg s$^{-1}$ with method 2. 
The $1 \sigma$ uncertainties on luminosities from method 2 were evaluated from the luminosity distributions obtained from the Monte Carlo sampling of the $EMD(T)$ and abundances parameter space. In the case of method 1, we assume an uncertainty on $L_{\rm X}$ equal to the range measured between the Feb 2021 observation and the high-state at the Aug 2021 epoch, which is 1.6 -- $2.4 \times 10^{30}$\, erg\,s$^{-1}$ in the 5--100\AA\ band, while the uncertainty on the EUV luminosity was computed by generating the spectra relative to the upper and lower boundaries of the EMD range.

In Fig.\ref{fig:xuvscale} we plotted the values derived with the two methods, and a few other benchmark stars employed in the past to calibrate the old scaling laws proposed by \citet{SF11} (SF11), \citet{Chadney+2015} (C15), \citet{King+2018} (K18), and the new scaling laws by \citet{Johnstone+2021} (J21) and by \citet{SF22} (SF22).
In Fig.\ref{fig:xuvscale} we do not show the C15 scaling law, because it was computed in a different EUV band (80--350\,\AA).

In the cases C15, K18 and J21, the original scaling laws employed X-ray and EUV fluxes at the stellar surface, hence we computed scaling laws in luminosity assuming two different stellar radii, 0.7\,$R_\odot$ and 1.3\,$R_\odot$, representative of stars ranging from an early M-type dwarf such as AU\,Mic to a pre-main-sequence solar mass star such as \target. In the case SF22, in Fig.\ref{fig:xuvscale} we show the confidence region at the 95\% level of the least squares linear regression, in the range of validity of this scaling law, and its extrapolation to the position of \target. 

Among the benchmark stars, selected to cover a wide range of activity levels, we included our Sun, with flux ranges derived from \citet[]{Johnstone+2021} and based on observations with the TIMED/SEE mission. As intermediate-activity stars, we show the cases of HD\,189733 (SF1, and \citealt{Bourrier+2020}) and $\epsilon$\,Eri (SF11, C15 and K18). For the latter, we adopted the X-ray luminosity range derived by \citet{Coffaro+2020}, because the EUV measurement is not simultaneous. Finally, as a prototype of a young high-activity red dwarf we selected AU\,Mic (K18 and C15), with X-ray and EUV luminosity ranges taking into account source variability and/or measurement uncertainties. We stress that the most recent scaling laws (J21 and SF22) are based on a few tens of stars observed in X-rays and UV or EUV wavelengths at different epochs.

\section{Discussion and conclusions} \label{sec:discuss}

In the general case of low- or intermediate activity stars, such as our Sun or $\epsilon$\,Eri, the EUV and X-ray stellar emission is characterized by substantial variability on both short and long time scales, due to phenomena ranging from rotational modulation to magnetic cycles. Hence, in order to improve our capability to predict the high-energy irradiation of exoplanets we need coordinated observation campaigns in different bands, as performed for \target. 

Our analysis of the chromospheric and coronal emission of \target\ and its variability on time scales ranging from a few hours to several years suggests instead that this PMS star has a fairly high and steady activity level, typical of coronal sources in the saturated regime \citep{Pizz03}. 
Comparison of available measurements obtained with ROSAT, Chandra, and XMM-Newton, showed variability of the X-ray luminosity within a factor $\sim 2$. A similar amplitude is indicated by our 3-year time series of the chromospheric \ion{Ca}{2} H\&K emission, and by comparison of the \ion{Si}{4} resonant doublet in our HST/COS G130M observation with that in a COS G160M spectrum taken ten months before. 
This result is consistent with the trend of decreasing amplitude of magnetic activity cycles with increasing stellar high-energy fluxes \citep{wargelin+2017,Coffaro+2022}.

In the present case, the lack of strong variability allowed us to employ HST/COS observations together with XMM-Newton spectroscopic data taken 6 months apart, with the primary goal to add a new benchmark point on the EUV vs.\ X-ray luminosity relationship. This point is representative of an active PMS solar-mass star.

To this aim, we have employed a classical approach, namely the reconstruction of the full plasma emission measure distribution from chromosphere to corona, based on accurate measurements of emission lines which form over a wide range of temperatures. The synthetic XUV spectrum allowed us to estimate the stellar EUV flux, which cannot be directly observed with any present space facility. As a side result, we have also obtained measurements of chemical abundances in the stellar outer atmosphere.

We have adopted different approaches and methodologies, which converge toward consistent and robust results, but which highlight also systematic uncertainties, unavoidable with current instrumentation and knowledge of atomic physics. In particular, we have obtained two predictions of the EUV luminosity (or flux) which differ by a factor 1.5, depending on the method and atomic database adopted. This uncertainty yields a mean EUV to X-ray luminosity ratio (or $F_{\rm EUV}/F_{\rm X}$) of $0.6 \pm 0.1$.

The old SF11 scaling law predicts for \target\ an EUV luminosity significantly higher than observed, while the K18 version yields a lower value. The new law proposed by J21 (steeper than K18) provides a better approximation to the \target\ benchmark position than the new formula by SF22, although shallower than SF11.
More precisely, the EUV luminosity computed with method 1 is about a factor 2.5 lower than the SF22 prediction, but the error bar overlaps with the 95\% confidence region of the SF22 scaling law (Fig.\ \ref{fig:xuvscale}), and the same occurs for the $F_{\rm EUV}/F_{\rm X}$ ratio. On the other hand, the measurements with method 2 result just a factor 1.3 lower than the J21 prediction, but again compatible within the $1\sigma$ uncertainties.
We recall that the full stellar samples employed to derive these scaling laws show a standard deviation of about a factor 3 with respect to the least squares analytic solutions. 

A refinement of the EUV vs. X-ray scaling law is beyond the scope of the present work. However, we recall a number of differences and critical issues, which may require future investigations. The SF22 solution is derived with a line-based approach, similar to that employed for \target. This approach allows the direct computation of EUV fluxes and luminosities in the full band 100--920\,\AA, but it is subject to the uncertainties in the reconstructed Emission Measure Distributions and on the atomic database employed for this aim. On the other hand, the J21 solution is based on broad-band (5.17--124\,\AA) X-ray fluxes measured with the ROSAT satellite, and EUV fluxes in the 100--360\,\AA\ band, computed by direct integration of spectra taken with the EUVE satellite. Extension to the full 100--920\,\AA\ is provided with a further scaling law which relates the EUVE fluxes to those in the 360--920\,\AA\ range, calibrated only on solar data. Moreover, the J21 solution relies on accurate knowledge of stellar radii, which are not always available.

Both solutions assume essentially that the X-ray to EUV scaling law is independent from mass and relates uniquely to stars with activity levels ranging from the quiet Sun to young PMS stars. While useful for further prediction of the high-energy irradiation of exoplanets, this hypothesis needs to be tested with further multi-wavelengths observations of targets with similar mass but different ages and activity levels.

In conclusion, we recall that different predictions of the stellar EUV flux and its long-term evolution affect the time scale of photoevaporation of planetary atmospheres and the possibility to reach stability or to lose them entirely (\citealt{Maggio+2022}), thus providing different forecasts to perform transmission and emission spectroscopy of exoplanets with JWST and the future Ariel mission.



\begin{acknowledgments}
We acknowledge financial contributions from the ASI-INAF agreement n.2018-16-HH.0 (THE StellaR PAth project), and from the ARIEL ASI-INAF agreement n.2021-5-HH.0. 
C.A.\ acknowledges funding from University of Palermo FFR-2023.
J.S.F.\ acknowledges financial support from the Agencia Estatal de Investigaci\'on of the Spanish MCIN, MCIN/AEI/10.13039/501100011033, through grant PID2019-109522GB-C51.
We also acknowledge partial support by the project HOT-ATMOS (PRIN INAF 2019). 

This work is based on observations obtained with \xmm, an ESA science mission with instruments and contributions directly funded by ESA Member States and NASA; also based on observations made with the NASA/ESA Hubble Space Telescope, obtained at the Space Telescope Science Institute, and on observations made with the Italian Telescopio Nazionale Galileo (TNG) operated on the island of La Palma by the Fundaci\'on Galileo Galilei of the INAF (Istituto Nazionale di Astrofisica) at the Spanish Observatorio del Roque de los Muchachos of the Instituto de Astrof\'isica de Canarias.
\end{acknowledgments}


\vspace{5mm}
\facilities{XMM (EPIC and RGS); HST (COS); TNG (HARPS-N)}

All the {\it HST} data used in this paper can be found in MAST: \dataset[10.17909/8sff-bk67]{http://dx.doi.org/10.17909/8sff-bk67}.


\software{SAS
(\citealt{SAS2004}, \url{www.cosmos.esa.int/web/xmm-newton/sas}),
XSPEC (\citealt{XSPEC1996}, \url{heasarc.gsfc.nasa.gov/xanadu/xspec}), PINTofALE (\citealt{Kashyap1998}, \url{http://hea-www.harvard.edu/PINTofALE/},
Python (v3.7) with the libraries numpy and matplotlib, ISIS (\url{space.mit.edu/CXC/isis})
%
}




\appendix
\section{Optical depth effects}
\label{Appendix}

All the UV and X-ray analysis presented in the paper is based on the assumption that the plasma responsible for the UV and X-ray emission from \target\ is collisionally-excited and optically-thin. For increasing optical depth, opacity effects first appear in strong resonance lines, because at those wavelengths the absorption coefficient (and hence the optical depth) has local maxima. In these cases, since the probability of photons absorption and re-emission in other directions is non-negligible, the line intensity is reduced if compared to the corresponding optically-thin case. This phenomenon is called resonant scattering. To check whether resonant scattering is present, flux ratios of lines corresponding to different absorption coefficients have to be compared with the ratio expected in case of collisionally-excited optically-thin emission.

\begin{figure}
    \centering
    \resizebox{0.4\textwidth}{!}{\includegraphics{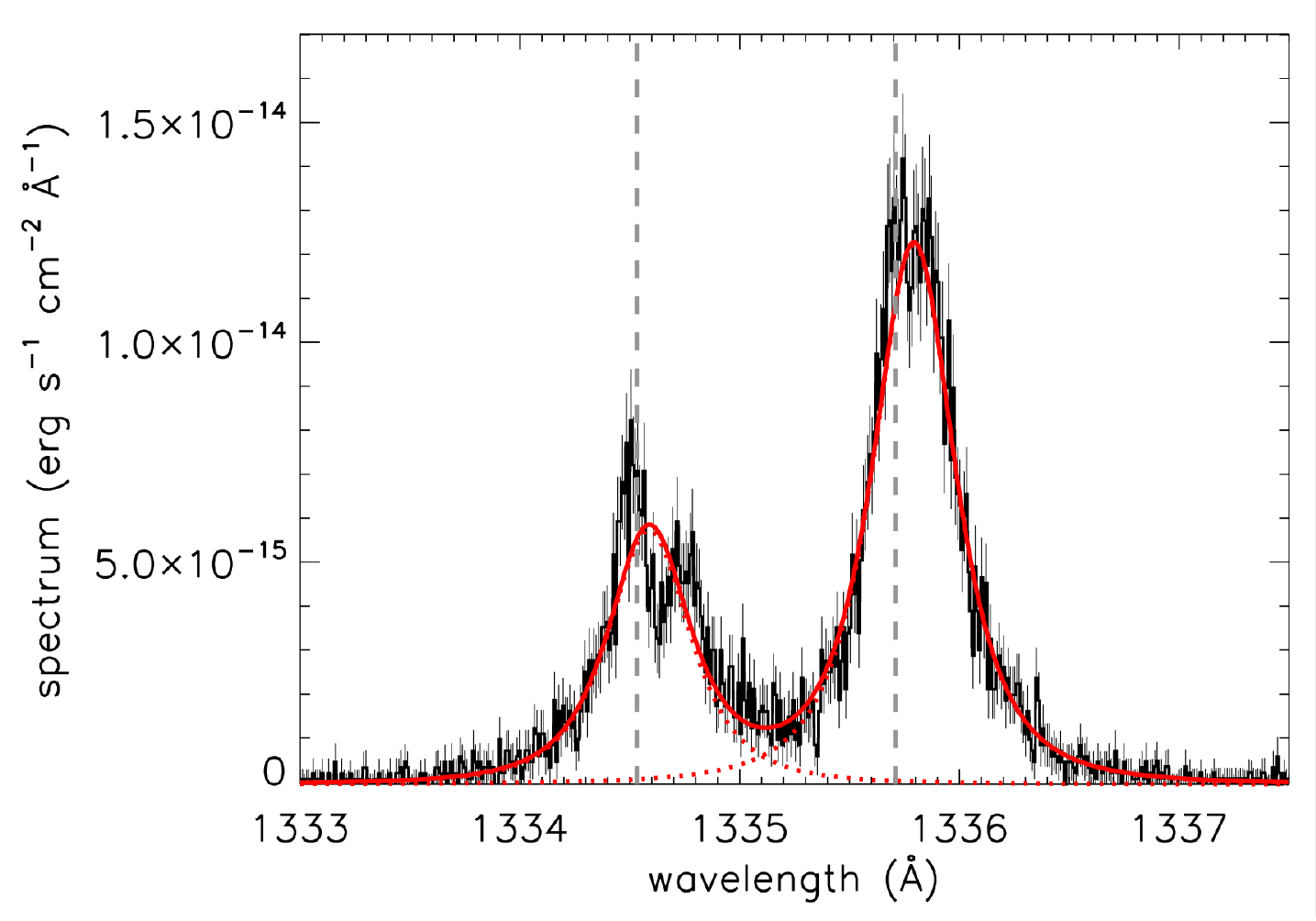}}
    \caption{Observed line profiles, in black, of two \ion{C}{2} lines in the COS spectrum. The best-fit function used to infer line fluxes is shown in red (solid line marks the total best fit function, dotted lines the contributions of individual emission lines).}
    \label{fig:opticaldepth}
\end{figure}

\begin{figure}
    \centering
    \resizebox{0.48\textwidth}{!}{\includegraphics[trim=2.8cm 14.0cm 0.0cm 3cm, clip]{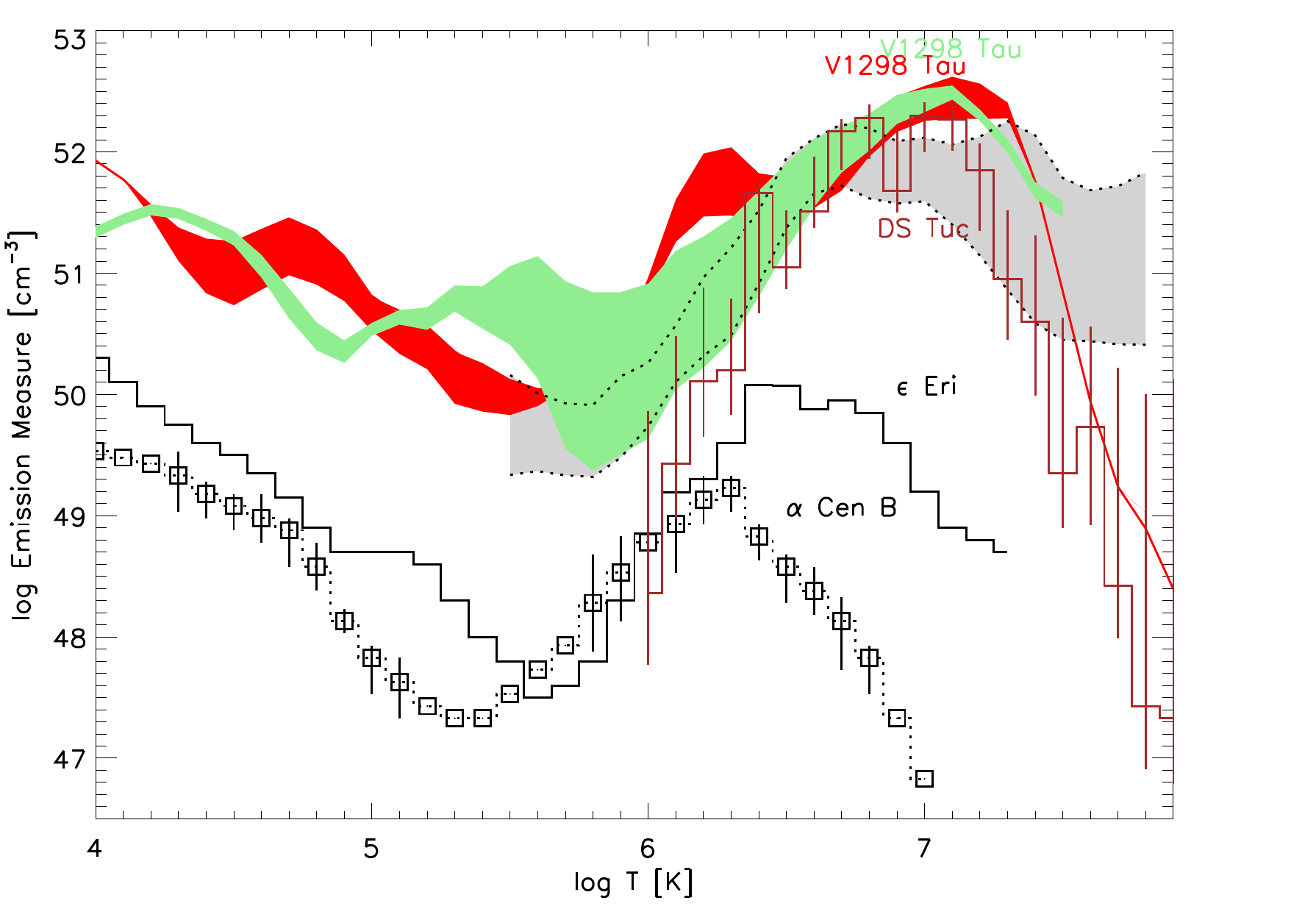}}
    \caption{Comparison of plasma Emission Measure Distributions vs.\ temperature for \target\ and three other G--K stars with different activity levels (see text). A polynomial smoothing (order 2, 6 bins width) was applied to the low and high $1 \sigma$ boundaries of the two EMD solutions presented in Sect.\ \ref{sect:dem} (red band for method 1, green band for method 2). The gray band shows instead the range predicted by \cite{Wood+2018} for stars with surface X-ray fluxes between $10^7$ and $3 \times 10^7$\,erg s$^{-1}$ cm$^{-2}$.}
    \label{fig:dems}
\end{figure}

Optical depth effects, if present, are expected in chromospheric lines, because these atmospheric layers have densities higher than that of coronal structures. To check whether optical depth effects affect some chromospheric lines we inspected the Li-like and Na-like doublets of \ion{C}{4} (1548.19 and 1550.78\,\AA), \ion{N}{5} (1238.82 and 1242.81\,\AA), and \ion{Si}{4} (1393.76 and 1402.77\,\AA). In the optically-thin collisionally-excited regime, the predicted ratio of the two lines of each doublet is 2:1. These ratios do not depend on temperature or abundances, and have a negligible dependence on interstellar absorption. Therefore the comparison between observed and predicted ratios can be performed irrespective of any source modeling. 

The unabsorbed observed ratios of these three doublets are $1.85\pm0.09$, $1.83\pm0.14$, and $1.76\pm0.10$, respectively. These values, slightly lower than those predicted, indicate that for the strongest lines of the inspected doublets, opacity effects are present but modest. This conclusion is further supported by the line profile observed in a few cases of strong UV lines. The clearest case is the profile of the two \ion{C}{2} at $\sim1335$\,\AA\, (fig~\ref{fig:opticaldepth}). Both the profiles clearly indicate a modest but clear line quenching near the peak. We exclude that this kind of line profile can be due to other effects, like for instance different velocity components, because, in that case, the same profile should characterize all the lines originating from the same plasma component. Notice however that measuring the fluxes of these lines using anyhow a single Gaussian encompassing the pairs of peaks, allows us to partially reconstruct the missing flux near the line peak.

In all cases (the inspected line doublets whose ratio marginally deviates from the optically-thin case, and the line profiles showing flux quenching) we are possibly underestimating the rate of emitted photons by a factor of $\sim10\%$. This factor is anyhow comparable with the uncertainties that characterize the derived EMD.

\section{\target\ in comparison}
\label{appendix2}

In order to put \target\ in the wider context of coronal X-ray sources, in Fig.\ \ref{fig:dems} we compare the Emission Measure Distribution (EMD) vs.\ temperature of \target, derived with the two methods described in Sect.\ \ref{sect:dem}, with three other G--K stars with different activity levels and corresponding high-energy fluxes. In particular, we selected $\alpha$\,Cen B (K1V) as a prototype of a low-activity star \citep{SF11}, $\epsilon$\,Eri (K2V) as an example of an intermediate-activity coronal source \citep{SanzForcada+2003}, and DS Tuc A (G6V) as a high-activity object \citep{Pillitteri+2022}. The latter is also a young planet-hosting star, about 40\,Myr old, for which a study of the photo-evaporation of the planetary atmosphere was presented by \cite{Benatti+2021}, and strong flares were detected by XMM-Newton \citep{Pillitteri+2022}, but we show just the EMD of the quiescent corona here for comparison.

The shift of the peak emission measure toward higher coronal temperatures, from $\sim 2$\,MK to 10\,MK for increasing activity level, was already noted by several authors in the past (see for example \citealt{Scelsi+2005}). A similar trend is visible for the temperature of the minimum emission measures, but the variation is within a factor 3, from $10^{5.3}$\,K to $10^{5.8}$\,K. In the chromospheric region, from $10^4$\,K to the temperatures at the minimum, the EMDs can be approximated with a power law, $\propto T^{-\alpha}$, with $\alpha$ between 1.5 and 1.7,
for all stars. This similarity was already assessed by \cite{SF11}.

In the same figure, we show the EMD range predicted for very active stars according to \cite{Wood+2018}, who analyzed the X-ray spectra of 19 late-type dwarfs observed by Chandra, and showed a possible scaling of the EMDs with the surface X-ray flux. The coronal EMD of \target, for temperatures above 1\,MK, results in nice overlap with the range of values expected for stars having $\log F_{\rm x}$ between 7.0 and 7.5\,erg s$^{-1}$ cm$^{-2}$, i.e.\ those with the highest high-energy emission level in the study by \cite{Wood+2018}.


\bibliography{v1298}{}

\begin{thebibliography}{}
\expandafter\ifx\csname natexlab\endcsname\relax\def\natexlab#1{#1}\fi
\providecommand{\url}[1]{\href{#1}{#1}}
\providecommand{\dodoi}[1]{doi:~\href{http://doi.org/#1}{\nolinkurl{#1}}}
\providecommand{\doeprint}[1]{\href{http://ascl.net/#1}{\nolinkurl{http://ascl.net/#1}}}
\providecommand{\doarXiv}[1]{\href{https://arxiv.org/abs/#1}{\nolinkurl{https://arxiv.org/abs/#1}}}

\bibitem[{{Anders} \& {Grevesse}(1989)}]{AG1989}
{Anders}, E., \& {Grevesse}, N. 1989, \gca, 53, 197,
  \dodoi{10.1016/0016-7037(89)90286-X}

\bibitem[{{Arnaud}(1996)}]{XSPEC1996}
{Arnaud}, K.~A. 1996, in Astronomical Society of the Pacific Conference Series,
  Vol. 101, Astronomical Data Analysis Software and Systems V, ed. G.~H.
  {Jacoby} \& J.~{Barnes}, 17

\bibitem[{{Benatti} {et~al.}(2021){Benatti}, {Damasso}, {Borsa}, {Locci},
  {Pillitteri}, {Desidera}, {Maggio}, {Micela}, {Wolk}, {Claudi}, {Malavolta},
  \& {Modirrousta-Galian}}]{Benatti+2021}
{Benatti}, S., {Damasso}, M., {Borsa}, F., {et~al.} 2021, \aap, 650, A66,
  \dodoi{10.1051/0004-6361/202140416}

\bibitem[{{Bourrier} {et~al.}(2020){Bourrier}, {Wheatley}, {Lecavelier des
  Etangs}, {King}, {Louden}, {Ehrenreich}, {Fares}, {Helling}, {Llama},
  {Jardine}, \& {Vidotto}}]{Bourrier+2020}
{Bourrier}, V., {Wheatley}, P.~J., {Lecavelier des Etangs}, A., {et~al.} 2020,
  \mnras, 493, 559, \dodoi{10.1093/mnras/staa256}

\bibitem[{{Carleo} {et~al.}(2020){Carleo}, {Malavolta}, {Lanza}, {Damasso},
  {Desidera}, {Borsa}, {Mallonn}, {Pinamonti}, {Gratton}, {Alei}, {Benatti},
  {Mancini}, {Maldonado}, {Biazzo}, {Esposito}, {Frustagli},
  {Gonz{\'a}lez-{\'A}lvarez}, {Micela}, {Scandariato}, {Sozzetti}, {Affer},
  {Bignamini}, {Bonomo}, {Claudi}, {Cosentino}, {Covino}, {Fiorenzano},
  {Giacobbe}, {Harutyunyan}, {Leto}, {Maggio}, {Molinari}, {Nascimbeni},
  {Pagano}, {Pedani}, {Piotto}, {Poretti}, {Rainer}, {Redfield}, {Baffa},
  {Baruffolo}, {Buchschacher}, {Billotti}, {Cecconi}, {Falcini}, {Fantinel},
  {Fini}, {Galli}, {Ghedina}, {Ghinassi}, {Giani}, {Gonzalez}, {Gonzalez},
  {Guerra}, {Hernandez Diaz}, {Hernandez}, {Iuzzolino}, {Lodi}, {Oliva},
  {Origlia}, {Perez Ventura}, {Puglisi}, {Riverol}, {Riverol}, {San Juan},
  {Sanna}, {Scuderi}, {Seemann}, {Sozzi}, \& {Tozzi}}]{Carleo2020}
{Carleo}, I., {Malavolta}, L., {Lanza}, A.~F., {et~al.} 2020, \aap, 638, A5,
  \dodoi{10.1051/0004-6361/201937369}

\bibitem[{{Cecchi-Pestellini} {et~al.}(2006){Cecchi-Pestellini}, {Ciaravella},
  \& {Micela}}]{ChecchiPestellini+2006}
{Cecchi-Pestellini}, C., {Ciaravella}, A., \& {Micela}, G. 2006, \aap, 458,
  L13, \dodoi{10.1051/0004-6361:20066093}

\bibitem[{{Chadney} {et~al.}(2015){Chadney}, {Galand}, {Unruh}, {Koskinen}, \&
  {Sanz-Forcada}}]{Chadney+2015}
{Chadney}, J.~M., {Galand}, M., {Unruh}, Y.~C., {Koskinen}, T.~T., \&
  {Sanz-Forcada}, J. 2015, \icarus, 250, 357,
  \dodoi{10.1016/j.icarus.2014.12.012}

\bibitem[{{Coffaro} {et~al.}(2022){Coffaro}, {Stelzer}, \&
  {Orlando}}]{Coffaro+2022}
{Coffaro}, M., {Stelzer}, B., \& {Orlando}, S. 2022, Astronomische Nachrichten,
  343, e10066, \dodoi{10.1002/asna.20210066}

\bibitem[{{Coffaro} {et~al.}(2020){Coffaro}, {Stelzer}, {Orlando}, {Hall},
  {Metcalfe}, {Wolter}, {Mittag}, {Sanz-Forcada}, {Schneider}, \&
  {Ducci}}]{Coffaro+2020}
{Coffaro}, M., {Stelzer}, B., {Orlando}, S., {et~al.} 2020, \aap, 636, A49,
  \dodoi{10.1051/0004-6361/201936479}

\bibitem[{{Cohen} {et~al.}(2014){Cohen}, {Drake}, {Glocer}, {Garraffo},
  {Poppenhaeger}, {Bell}, {Ridley}, \& {Gombosi}}]{Cohen+2014}
{Cohen}, O., {Drake}, J.~J., {Glocer}, A., {et~al.} 2014, \apj, 790, 57,
  \dodoi{10.1088/0004-637X/790/1/57}

\bibitem[{{Cosentino} {et~al.}(2014){Cosentino}, {Lovis}, {Pepe}, {Collier
  Cameron}, {Latham}, {Molinari}, {Udry}, {Bezawada}, {Buchschacher},
  {Figueira}, {Fleury}, {Ghedina}, {Glenday}, {Gonzalez}, {Guerra}, {Henry},
  {Hughes}, {Maire}, {Motalebi}, \& {Phillips}}]{Cosentino2014}
{Cosentino}, R., {Lovis}, C., {Pepe}, F., {et~al.} 2014, in Society of
  Photo-Optical Instrumentation Engineers (SPIE) Conference Series, Vol. 9147,
  Ground-based and Airborne Instrumentation for Astronomy V, ed. S.~K.
  {Ramsay}, I.~S. {McLean}, \& H.~{Takami}, 91478C, \dodoi{10.1117/12.2055813}

\bibitem[{{David} {et~al.}(2019){David}, {Petigura}, {Luger}, {Foreman-Mackey},
  {Livingston}, {Mamajek}, \& {Hillenbrand}}]{David+2019b}
{David}, T.~J., {Petigura}, E.~A., {Luger}, R., {et~al.} 2019, \apjl, 885, L12,
  \dodoi{10.3847/2041-8213/ab4c99}

\bibitem[{{Di Maio} {et~al.}(2020){Di Maio}, {Argiroffi}, {Micela}, {Benatti},
  {Lanza}, {Scandariato}, {Maldonado}, {Maggio}, {Affer}, \&
  {Claudi}}]{dimaio2020}
{Di Maio}, C., {Argiroffi}, C., {Micela}, G., {et~al.} 2020, \aap, 642, A53,
  \dodoi{10.1051/0004-6361/202038011}

\bibitem[{{Fitzpatrick}(1999)}]{Fitzpatrick1999}
{Fitzpatrick}, E.~L. 1999, \pasp, 111, 63, \dodoi{10.1086/316293}

\bibitem[{{Foster} {et~al.}(2012){Foster}, {Ji}, {Smith}, \&
  {Brickhouse}}]{aped}
{Foster}, A.~R., {Ji}, L., {Smith}, R.~K., \& {Brickhouse}, N.~S. 2012, \apj,
  756, 128, \dodoi{10.1088/0004-637X/756/2/128}

\bibitem[{{Fuhrmeister} {et~al.}(2022){Fuhrmeister}, {Czesla}, {Robrade},
  {Gonz{\'a}lez-P{\'e}rez}, {Schneider}, {Mittag}, \&
  {Schmitt}}]{Fuhrmeister+2022}
{Fuhrmeister}, B., {Czesla}, S., {Robrade}, J., {et~al.} 2022, \aap, 661, A24,
  \dodoi{10.1051/0004-6361/202141020}

\bibitem[{{Fulton} \& {Petigura}(2018)}]{Fulton+Petigura2018}
{Fulton}, B.~J., \& {Petigura}, E.~A. 2018, \aj, 156, 264,
  \dodoi{10.3847/1538-3881/aae828}

\bibitem[{{Fulton} {et~al.}(2017){Fulton}, {Petigura}, {Howard}, {Isaacson},
  {Marcy}, {Cargile}, {Hebb}, {Weiss}, {Johnson}, {Morton}, {Sinukoff},
  {Crossfield}, \& {Hirsch}}]{Fulton+2017}
{Fulton}, B.~J., {Petigura}, E.~A., {Howard}, A.~W., {et~al.} 2017, \aj, 154,
  109, \dodoi{10.3847/1538-3881/aa80eb}

\bibitem[{{Gabriel} {et~al.}(2004){Gabriel}, {Denby}, {Fyfe}, {Hoar}, {Ibarra},
  {Ojero}, {Osborne}, {Saxton}, {Lammers}, \& {Vacanti}}]{SAS2004}
{Gabriel}, C., {Denby}, M., {Fyfe}, D.~J., {et~al.} 2004, in Astronomical
  Society of the Pacific Conference Series, Vol. 314, Astronomical Data
  Analysis Software and Systems (ADASS) XIII, ed. F.~{Ochsenbein}, M.~G.
  {Allen}, \& D.~{Egret}, 759

\bibitem[{{Gaia Collaboration}(2020)}]{GaiaDR3}
{Gaia Collaboration}. 2020, VizieR Online Data Catalog, I/350

\bibitem[{{Gomes da Silva} {et~al.}(2018){Gomes da Silva}, {Figueira},
  {Santos}, \& {Faria}}]{gomesdasilva18}
{Gomes da Silva}, J., {Figueira}, P., {Santos}, N., \& {Faria}, J. 2018, The
  Journal of Open Source Software, 3, 667, \dodoi{10.21105/joss.00667}

\bibitem[{{Johnstone} {et~al.}(2021){Johnstone}, {Bartel}, \&
  {G{\"u}del}}]{Johnstone+2021}
{Johnstone}, C.~P., {Bartel}, M., \& {G{\"u}del}, M. 2021, \aap, 649, A96,
  \dodoi{10.1051/0004-6361/202038407}

\bibitem[{{Kashyap} \& {Drake}(1998)}]{Kashyap1998}
{Kashyap}, V., \& {Drake}, J.~J. 1998, \apj, 503, 450, \dodoi{10.1086/305964}

\bibitem[{{Khodachenko} {et~al.}(2007){Khodachenko}, {Ribas}, {Lammer},
  {Grie{\ss}meier}, {Leitner}, {Selsis}, {Eiroa}, {Hanslmeier}, {Biernat},
  {Farrugia}, \& {Rucker}}]{Khodachenko+2007}
{Khodachenko}, M.~L., {Ribas}, I., {Lammer}, H., {et~al.} 2007, Astrobiology,
  7, 167, \dodoi{10.1089/ast.2006.0127}

\bibitem[{{King} {et~al.}(2018){King}, {Wheatley}, {Salz}, {Bourrier},
  {Czesla}, {Ehrenreich}, {Kirk}, {Lecavelier des Etangs}, {Louden}, {Schmitt},
  \& {Schneider}}]{King+2018}
{King}, G.~W., {Wheatley}, P.~J., {Salz}, M., {et~al.} 2018, \mnras, 478, 1193,
  \dodoi{10.1093/mnras/sty1110}

\bibitem[{{Lammer} {et~al.}(2003){Lammer}, {Selsis}, {Ribas}, {Guinan},
  {Bauer}, \& {Weiss}}]{Lammer+2003}
{Lammer}, H., {Selsis}, F., {Ribas}, I., {et~al.} 2003, \apjl, 598, L121,
  \dodoi{10.1086/380815}

\bibitem[{{Lammer} {et~al.}(2007){Lammer}, {Lichtenegger}, {Kulikov},
  {Grie{\ss}meier}, {Terada}, {Erkaev}, {Biernat}, {Khodachenko}, {Ribas},
  {Penz}, \& {Selsis}}]{Lammer+2007}
{Lammer}, H., {Lichtenegger}, H. I.~M., {Kulikov}, Y.~N., {et~al.} 2007,
  Astrobiology, 7, 185, \dodoi{10.1089/ast.2006.0128}

\bibitem[{{Locci} {et~al.}(2019){Locci}, {Cecchi-Pestellini}, \&
  {Micela}}]{Locci+2019}
{Locci}, D., {Cecchi-Pestellini}, C., \& {Micela}, G. 2019, \aap, 624, A101,
  \dodoi{10.1051/0004-6361/201834491}

\bibitem[{{Lopez} \& {Fortney}(2013)}]{Lopez+Fortney2013}
{Lopez}, E.~D., \& {Fortney}, J.~J. 2013, \apj, 776, 2,
  \dodoi{10.1088/0004-637X/776/1/2}

\bibitem[{{Lovis} {et~al.}(2011){Lovis}, {Dumusque}, {Santos}, {Bouchy},
  {Mayor}, {Pepe}, {Queloz}, {S{\'e}gransan}, \& {Udry}}]{2011arXiv1107.5325L}
{Lovis}, C., {Dumusque}, X., {Santos}, N.~C., {et~al.} 2011, arXiv e-prints,
  arXiv:1107.5325.
\newblock \doarXiv{1107.5325}

\bibitem[{{Maggio} {et~al.}(2007){Maggio}, {Flaccomio}, {Favata}, {Micela},
  {Sciortino}, {Feigelson}, \& {Getman}}]{Maggio+2007}
{Maggio}, A., {Flaccomio}, E., {Favata}, F., {et~al.} 2007, \apj, 660, 1462,
  \dodoi{10.1086/513088}

\bibitem[{{Maggio} {et~al.}(2022){Maggio}, {Locci}, {Pillitteri}, {Benatti},
  {Claudi}, {Desidera}, {Micela}, {Damasso}, {Sozzetti}, \& {Suarez
  Mascare{\~n}o}}]{Maggio+2022}
{Maggio}, A., {Locci}, D., {Pillitteri}, I., {et~al.} 2022, \apj, 925, 172,
  \dodoi{10.3847/1538-4357/ac4040}

\bibitem[{{Mittag} {et~al.}(2013){Mittag}, {Schmitt}, \&
  {Schr{\"o}der}}]{Mittag+2013}
{Mittag}, M., {Schmitt}, J.~H.~M.~M., \& {Schr{\"o}der}, K.~P. 2013, \aap, 549,
  A117, \dodoi{10.1051/0004-6361/201219868}

\bibitem[{{Oh} {et~al.}(2017){Oh}, {Price-Whelan}, {Hogg}, {Morton}, \&
  {Spergel}}]{Oh+2017}
{Oh}, S., {Price-Whelan}, A.~M., {Hogg}, D.~W., {Morton}, T.~D., \& {Spergel},
  D.~N. 2017, \aj, 153, 257, \dodoi{10.3847/1538-3881/aa6ffd}

\bibitem[{{Owen} \& {Lai}(2018)}]{Owen+Lai2018}
{Owen}, J.~E., \& {Lai}, D. 2018, \mnras, 479, 5012,
  \dodoi{10.1093/mnras/sty1760}

\bibitem[{{Owen} \& {Wu}(2013)}]{Ower+Wu2013}
{Owen}, J.~E., \& {Wu}, Y. 2013, \apj, 775, 105,
  \dodoi{10.1088/0004-637X/775/2/105}

\bibitem[{{Owen} \& {Wu}(2017)}]{Owen+Wu2017}
---. 2017, \apj, 847, 29, \dodoi{10.3847/1538-4357/aa890a}

\bibitem[{{Pillitteri} {et~al.}(2022){Pillitteri}, {Argiroffi}, {Maggio},
  {Micela}, {Benatti}, {Reale}, {Colombo}, \& {Wolk}}]{Pillitteri+2022}
{Pillitteri}, I., {Argiroffi}, C., {Maggio}, A., {et~al.} 2022, \aap, 666,
  A198, \dodoi{10.1051/0004-6361/202244268}

\bibitem[{{Pizzolato} {et~al.}(2003){Pizzolato}, {Maggio}, {Micela},
  {Sciortino}, \& {Ventura}}]{Pizz03}
{Pizzolato}, N., {Maggio}, A., {Micela}, G., {Sciortino}, S., \& {Ventura}, P.
  2003, \aap, 397, 147, \dodoi{10.1051/0004-6361:20021560}

\bibitem[{{Poppenhaeger} {et~al.}(2021){Poppenhaeger}, {Ketzer}, \&
  {Mallonn}}]{Poppenhaeger+2020}
{Poppenhaeger}, K., {Ketzer}, L., \& {Mallonn}, M. 2021, \mnras, 500, 4560,
  \dodoi{10.1093/mnras/staa1462}

\bibitem[{{Raymond}(1988)}]{raym88}
{Raymond}, J.~C. 1988, in NATO ASIC Proc. 249: Hot Thin Plasmas in
  Astrophysics, ed. R.~Pallavicini, 3+.
\newblock
  \url{http://adsabs.harvard.edu/cgi-bin/nph-bib_query?bibcode=1988htpa.conf....3R&db_key=AST}

\bibitem[{{Sanz-Forcada} {et~al.}(2003{\natexlab{a}}){Sanz-Forcada},
  {Brickhouse}, \& {Dupree}}]{SanzForcada+2003}
{Sanz-Forcada}, J., {Brickhouse}, N.~S., \& {Dupree}, A.~K. 2003{\natexlab{a}},
  \apjs, 145, 147, \dodoi{10.1086/345815}

\bibitem[{{Sanz-Forcada} {et~al.}(2004){Sanz-Forcada}, {Favata}, \&
  {Micela}}]{SanzForcada+2004}
{Sanz-Forcada}, J., {Favata}, F., \& {Micela}, G. 2004, \aap, 416, 281,
  \dodoi{10.1051/0004-6361:20034466}

\bibitem[{{Sanz-Forcada} {et~al.}(2022){Sanz-Forcada}, {L{\'o}pez-Puertas},
  {Nortmann}, \& {Lamp{\'o}n}}]{SF22}
{Sanz-Forcada}, J., {L{\'o}pez-Puertas}, M., {Nortmann}, L., \& {Lamp{\'o}n},
  M. 2022, 21st Cambridge Workshop on Cool Stars, Stellar Systems, and the Sun,
  \dodoi{10.5281/zenodo.7561725}

\bibitem[{{Sanz-Forcada} {et~al.}(2003{\natexlab{b}}){Sanz-Forcada}, {Maggio},
  \& {Micela}}]{san03}
{Sanz-Forcada}, J., {Maggio}, A., \& {Micela}, G. 2003{\natexlab{b}}, \aap,
  408, 1087, \dodoi{10.1051/0004-6361:20031025}

\bibitem[{{Sanz-Forcada} {et~al.}(2011){Sanz-Forcada}, {Micela}, {Ribas},
  {Pollock}, {Eiroa}, {Velasco}, {Solano}, \& {Garc{\'\i}a-{\'A}lvarez}}]{SF11}
{Sanz-Forcada}, J., {Micela}, G., {Ribas}, I., {et~al.} 2011, \aap, 532, A6,
  \dodoi{10.1051/0004-6361/201116594}

\bibitem[{{Scelsi} {et~al.}(2007){Scelsi}, {Maggio}, {Micela}, {Briggs}, \&
  {G{\"u}del}}]{Scelsi+2007}
{Scelsi}, L., {Maggio}, A., {Micela}, G., {Briggs}, K., \& {G{\"u}del}, M.
  2007, \aap, 473, 589, \dodoi{10.1051/0004-6361:20077792}

\bibitem[{{Scelsi} {et~al.}(2005){Scelsi}, {Maggio}, {Peres}, \&
  {Pallavicini}}]{Scelsi+2005}
{Scelsi}, L., {Maggio}, A., {Peres}, G., \& {Pallavicini}, R. 2005, \aap, 432,
  671, \dodoi{10.1051/0004-6361:20041739}

\bibitem[{{Suarez Mascare{\~n}o} {et~al.}(2021){Suarez Mascare{\~n}o},
  {Damasso}, {Lodieu}, {Sozzetti}, {Bejar}, {Benatti}, {Zapatero Osorio},
  {Micela}, {Rebolo}, {Desidera}, {Murgas}, {Claudi}, {Gonzalez Hernandez},
  {Malavolta}, {del Burgo}, {D'Orazi}, {Amado}, {Locci}, {Tabernero},
  {Marzari}, {Aguado}, {Turrini}, {Cardona Guillen}, {Toledo Padron}, {Maggio},
  {Bauer}, {Caballero}, {Chinchilla}, {Esparza-Borges}, {Gonzalez-Alvarez},
  {Granzer}, {Luque}, {Martin}, {Nowak}, {Oshagh}, {Palle}, {Parviainen},
  {Quirrenbach}, {Reiners}, {Ribas}, {Strassmeier}, {Weber}, \&
  {Mallonn}}]{paperI+2021}
{Suarez Mascare{\~n}o}, A., {Damasso}, M., {Lodieu}, N., {et~al.} 2021, Nature
  Astronomy, in press, \dodoi{10.1038/s41550-021-01533-7}

\bibitem[{{Su{\'a}rez Mascare{\~n}o} {et~al.}(2021){Su{\'a}rez Mascare{\~n}o},
  {Damasso}, {Lodieu}, {Sozzetti}, {B{\'e}jar}, {Benatti}, {Zapatero Osorio},
  {Micela}, {Rebolo}, {Desidera}, {Murgas}, {Claudi}, {Gonz{\'a}lez
  Hern{\'a}ndez}, {Malavolta}, {del Burgo}, {D'Orazi}, {Amado}, {Locci},
  {Tabernero}, {Marzari}, {Aguado}, {Turrini}, {Cardona Guill{\'e}n},
  {Toledo-Padr{\'o}n}, {Maggio}, {Aceituno}, {Bauer}, {Caballero},
  {Chinchilla}, {Esparza-Borges}, {Gonz{\'a}lez-{\'A}lvarez}, {Granzer},
  {Luque}, {Mart{\'\i}n}, {Nowak}, {Oshagh}, {Pall{\'e}}, {Parviainen},
  {Quirrenbach}, {Reiners}, {Ribas}, {Strassmeier}, {Weber}, \&
  {Mallonn}}]{natpaper}
{Su{\'a}rez Mascare{\~n}o}, A., {Damasso}, M., {Lodieu}, N., {et~al.} 2021,
  arXiv e-prints, arXiv:2111.09193.
\newblock \doarXiv{2111.09193}

\bibitem[{{Wargelin} {et~al.}(2017){Wargelin}, {Saar}, {Pojma{\'n}ski},
  {Drake}, \& {Kashyap}}]{wargelin+2017}
{Wargelin}, B.~J., {Saar}, S.~H., {Pojma{\'n}ski}, G., {Drake}, J.~J., \&
  {Kashyap}, V.~L. 2017, \mnras, 464, 3281, \dodoi{10.1093/mnras/stw2570}

\bibitem[{{Wood} {et~al.}(2018){Wood}, {Laming}, {Warren}, \&
  {Poppenhaeger}}]{Wood+2018}
{Wood}, B.~E., {Laming}, J.~M., {Warren}, H.~P., \& {Poppenhaeger}, K. 2018,
  \apj, 862, 66, \dodoi{10.3847/1538-4357/aaccf6}

\end{thebibliography}
\bibliographystyle{aasjournal}



\end{document}